\documentclass[12pt,preprint]{aastex}
\usepackage{epstopdf}
\usepackage{graphicx}
\usepackage{amsmath,amsfonts}
\shortauthors{GWINN ET AL.}
\shorttitle{NOISE IN CROSS-POWER SPECTRA}
\begin{document}
\input epsf

\title{Noise in the Cross-Power Spectrum of the Vela Pulsar}
\author{C. R. Gwinn, M. D. Johnson}
\affil{Department of Physics, University of California, Santa Barbara, California 93106}
\email{cgwinn@physics.ucsb.edu} 

\author{J. E. Reynolds, D. L. Jauncey\altaffilmark{1}, and A. K. Tzioumis}
\affil{Australia Telescope National Facility, CSIRO, P.O. Box 76, Epping, NSW 1710, Australia}

\author{S. Dougherty, B. Carlson, and D. Del Rizzo}
\affil{National Research Council of Canada, Herzberg Institute for Astrophysics, Dominion Radio Astrophysical Observatory, PO Box 248, Penticton, BC, V2A 6J9, Canada}

\author{H.Hirabayashi, H. Kobayashi\altaffilmark{2}, Y. Murata, and P. G. Edwards\altaffilmark{3}}
\affil{The Institute of Space and Astronautical Science, Japan Aerospace Exploration Agency, 3-1-1 Yoshinodai, Chuou-ku Sagamihara, Kanagawa 252-5210, Japan}

\author{J. F. H. Quick and C. S. Flanagan}
\affil{Hartebeesthoek Radio Astronomy Observatory, P.O. Box 443, Krugersdorp 1740, South Africa}

\author{P. M. McCulloch}
\affil{School of Mathematics and Physics, University of Tasmania, Private Bag 37, Hobart, TAS 7001, Australia}
\altaffiltext{1}{Mount Stromlo Observatory, Cotter Road Weston ACT 2611 Australia}
\altaffiltext{2}{Present address: National Astronomical Observatory of Japan, 2-21-1 Osawa, Mitaka, Tokyo, 181-8588, Japan}
\altaffiltext{3}{Present address: Australia Telescope National Facility, CSIRO, P.O. Box 76, Epping, NSW 1710, Australia}
\vskip 1 truein
\begin{abstract}

We compare the noise in interferometric measurements of the Vela pulsar from ground- and space-based antennas with theoretical predictions. 
The noise depends on both the flux density and the interferometric phase of the source.
Because the Vela pulsar is bright and scintillating, these comparisons extend into both the low and high signal-to-noise regimes. 
Furthermore, our diversity of baselines explores the full range of variation in interferometric phase. 
We find excellent agreement between theoretical expectations and our estimates of noise among samples within the characteristic scintillation scales. 
Namely, the noise is drawn from an elliptical Gaussian distribution in the complex plane, centered on the signal. The major axis, aligned with the signal phase, varies quadratically with the signal, while the minor axis, at quadrature, varies with
the same linear coefficients. For weak signal, the noise approaches a circular Gaussian distribution. Both the variance and covariance of the noise are also affected by artifacts of digitization and correlation. In particular, we show that gating introduces correlations between nearby spectral channels.

\end{abstract}

\keywords{methods: data analysis -- techniques -- stars: pulsars -- pulsars: individual: Vela pulsar -- interstellar scattering}

\section{INTRODUCTION}

Radioastronomical observations yield a deterministic part, the signal; and a random part, noise
\citep{tms86}.
Consequently, the signal-to-noise ratio, the magnitude of the deterministic part divided by the standard deviation of the random part, characterizes them.
An understanding of the noise is fundamentally important because it provides a measure of the possibility of detecting a weak signal, and of the reliability
of the measurements of a detected signal. 
The noise is particularly important in situations where the signal varies, because the noise can mimic the signal.
%In situations where the signal varies, the noise is important because it 
%can mimic the signal.

Noise includes background noise from the instrument and sky. However, because all radioastronomical signals are noiselike 
(with one possible exception: \citet{jen01,Smi03}), they also contribute \emph{self-noise} \citep{kul89,ana91,viv91,mcc93,gwi06,gwi11a}. 
%All radioastronomical signals are noiselike 
%(with the exception of pulsar signals under certain circumstances: \citet{jen01}).
%The noiselike signal thus contributes to the noise in a measured spectrum.
As argued previously \citep{gwi06,gwi11a} and in Section\ \ref{sec:noise_dist_theory} below,
for interferometric visibility, the variance of the noise increases quadratically with the signal in phase with the signal,
and linearly with the signal at quadrature to the signal. The constant and linear coefficients in these two directions are equal.

We tested this picture for the cross-power spectrum of a scintillating source -- the Vela pulsar. 
These observations provided an ideal laboratory for such studies because
the pulsar varies greatly in flux density and interferometric phase with frequency and time,
because of interstellar scintillation \citep{des92}. 
Thus, each spectrum spanned many scintillation elements. 
Furthermore, the signal-to-noise ratio of the strongest spectral peaks can be high, even for short integrations,
and the source contributes significantly to system temperature, so that self-noise is important.

%In particular, the electric field of the source is drawn from a Gaussian distribution.
%Through the action of the central limit theorem,
%integration over frequency and time will drive the distribution of noise to a Gaussian distribution.
%Such a distribution is fully characterized by variances and covariances. 
%Because of self-noise, 
%the variance depends on signal amplitude; with different forms in phase with the signal and at quadrature \citep{gwi06,gwi11b}.

We compared predictions with observational estimates, 
formed by differencing samples close together in time 
and binning them according to their estimated average visibility.
This procedure provides for convenient visualization of the noise distribution.
Our investigation extends our previous results \citep{gwi11b} to the regime of high signal-to-noise ratio and large interferometric phase 
variations. 
Because of pulsar gating, 
the number of samples per integration time was small 
and depended on the number of gated pulses within the integration time.
From correlation functions, we determined the covariances of noise among spectral channels. 
Such covariances can be produced by quantization and
by correlator effects.  We discuss these effects and compare results with observations.

\subsection{Organization of this Paper}

Because the noise in the spectrum is drawn from a nearly Gaussian distribution with zero mean, variances and covariances
characterize it. 
In Section\ \ref{sec:theory_bckgnd}, 
we introduce the theoretical basis for noise in interferometric visibility and present the mathematical descriptions used in the paper.
In Section\ \ref{sec:observations}, we discuss our observations, correlation, and initial data processing. The remainder of the paper analyzes the noise in these observations. 
%We discuss our measurements of noise in the remainder of the paper.
In Section\ \ref{sec:noise_dist_obs}, we quantify the distribution of noise and the influence of a signal.  
In Section\ \ref{sec:correlator_effects}, we discuss quantization and correlator effects.  
In Section \ref{sec:conclusions}, we summarize our results and discuss implications for future 
observations and instruments.

\section{THEORETICAL BACKGROUND AND NOTATION}\label{sec:theory_bckgnd}

\subsection{Correlation Functions and Spectra}\label{sec:corrfunc_spectra}

Observations of electromagnetic radiation from astronomical sources measure and compare finite samples of electromagnetic fields.
We suppose that these are drawn from ensembles of statistically-identical measurements. 
Noise is the difference between the result of a measurement and the average of an infinite ensemble of such measurements.
The measurements, as well as their statistical averages, 
can be expressed either as spectra, varying with frequency, 
or as correlation functions, varying with lag in the Fourier-conjugate domain. 
In practice, the samples are digitized: quantized and time-sampled.
We introduce conventions for notation that help distinguish among these various domains.
The tilde denotes entities in the spectral domain,
indexed by the spectral channel as a subscript: for example, $\tilde r_k$.
Unaccented symbols designate the Fourier-conjugate domain of the correlation function, indexed by lag: $r_{\tau}$.
Angular brackets $\langle ...\rangle_{\rm n}$ denote a statistical average
over many realizations of the noiselike electric field,
with the scintillation spectrum held fixed.
The subscripted brackets $\langle ...\rangle_{\rm S}$ denote an average over many samples of the scintillation spectrum. 
These conventions are consistent with earlier work \citep{gwi04,gwi11a}.

As in earlier studies, we suppose that two antennas record time series of zero-mean, complex Gaussian random variables
\citep{gwi04,gwi06,gwi11a,gwi11b}. 
These time series can be regarded as the amplitude and phase of one polarization of electric field.
The antennas are $X$ and $Y$,
and the time series $x_{\ell}$ and $y_{\ell}$,
where $\ell$ indexes time. 
The ensemble-averaged cross-power spectrum $\tilde \rho_k$, and the ensemble-averaged autocorrelation spectrum $\tilde\alpha_k$, 
fully describe the spectral properties of these series.
At shorter wavelengths quantum-mechanical effects introduce shot noise, but the ensemble-averaged spectrum 
remains the same \citep{Zmu03}.
For convenience, we assume unit variances for the real and imaginary
parts.
%This assumption is in accord with quantization at sampling, which assigns particular values 
%according to a threshold \citep{gwi04}. 
%Spectral information resides in the covariances of samples.

An observer forms the cross-correlation function, $r_\tau$, of the time series $x_{\ell}$ and $y_{\ell}$.
The Fourier transform of $r_{\tau}$ gives the 
observed cross-power spectrum, $\tilde r_k$. This observed spectrum is an estimate of $\tilde\rho_k$, multiplied by a correlator-dependent gain factor (see Section \ref{sec::Correlator_Effects}). 
%usually multiplied by a gain factor $\Gamma$.
Similarly, autocorrelation of data at a single antenna yields an autocorrelation spectra $\tilde a_k$,
an estimate of $\tilde \alpha_k$.
The autocorrelation spectrum may include an offset, from the correlation of spectrally-flat background noise, 
as well as a gain factor.
In practice, individual measurements $\tilde r_k$ differ from 
the ``true'' spectrum $\tilde \rho_k$
by a random amount: the noise. 
If many samples of $\tilde r_k$ are averaged together, the noise for the average approaches a Gaussian distribution.

At lag $\tau$, the cross-correlation function $r_{\tau}$ is:
\begin{equation}\label{eq:rtau_def}
r_{\tau} = {\frac{1}{N_{\rm obs}-|\tau|}}\sum_{\ell=1}^{N_{\rm obs}-|\tau|} x_{\ell} y_{\ell+\tau}^* .
\end{equation}
This equation parallels Equation\ 6 of \citet{gwi04}; in particular,
$N_{\rm obs}$ is the number of samples.
However, we have reduced the definition of $r_{\tau}$ by a factor of 2, by changing the normalization of 
$x_{\ell}$ and $y_{\ell}$ so that $\langle |x_{\ell}|^2\rangle =\langle |y_{\ell}|^2\rangle =1$.
Note that
the sum here runs from 1 only to $N_{\rm obs}-|\tau|$,
so that large lags are averaged over fewer samples. 
If the lag $\tau < 0$, then the sum runs from $|\tau|+1$ to $N_{\rm obs}$. 
The finite spans of the time series prevent the correlation of all samples: for $\tau>0$, no $y_{\ell+\tau}$ exists for the last $\tau$ samples of $x_\ell$.  For $\tau<0$, no $x_{\ell}$ exists for the first $|\tau|$ samples of $y_{\ell+\tau}$. A pulsar gate, for example, can truncate the time series in this way.
Equation \ref{eq:rtau_def} describes the calculation of $r_{\tau}$ in many correlators of the "XF" design, where correlation "X" precedes Fourier transform "F". The S2 correlator at Penticton, used for the work described here, is of this design.
In contrast, our previous theoretical work assumed averaging of all lags $\tau$ over $N_{\rm obs}$ elements, 
because of the identification $x_{\ell - N_{\rm obs}} \equiv x_{\ell}$ for any $\ell$ \citep{gwi06,gwi11a}.
We call this the ``wrap'' assumption.
Most correlators of the alternative "FX" design, where Fourier transform precedes correlation, obey the "wrap" assumption. However, 
by zero-padding the time series they can emulate
the sum in Equation\ \ref{eq:rtau_def}.
We discuss the consequences of these two formulations for the cross-power spectrum
in Section\ \ref{sec:not_wrapping} below.

The statistical average of the cross-correlation function $r_{\tau}$ is 
$\langle r_{\tau}\rangle_{\rm n} = \rho_{\tau}$.  
This average runs over many realizations of the electric field, with uniform statistics.
Our observations are not stationary and do not approximate an ensemble average 
because
both the scintillation spectrum and the intrinsic flux density of the pulsar change with time.
However, we can regard each observation as being drawn from
an ensemble of observations of statistically-identical pulses at the same pulse phase and in the same scintillation state
\citep[see][]{gwi11b}.

The observed
cross-power spectrum is the Fourier transform of the cross-correlation function:
\begin{equation}
\tilde r_k=\sum_{\tau=-N}^{N-1} e^{i {\frac{2\pi}{2 N}} k \tau} r_\tau ,
\label{eq:Fourier_transf_xpower}
\end{equation}
where the number of spectral channels is $2N$.
The statistical average of $\tilde r_k$ is $\langle \tilde r_k \rangle_{\rm n} = \tilde \rho_k$.
For autocorrelation, the
average over $x_{\ell} x_{\ell+\tau}^*$ analogous to Equation\ \ref{eq:rtau_def}
forms the autocorrelation function $a_{\tau}$.
A statistical average of $a_{\tau}$ yields
$\langle a_{\tau} \rangle_{\rm n} = \alpha_{\tau}$.
The Fourier transform analogous to Equation\ \ref{eq:Fourier_transf_xpower} yields the autocorrelation spectrum
$\tilde a_k$, and an average over a statistical ensemble yields $\langle \tilde a_k \rangle_{\rm n} = \tilde \alpha_k$.

% \subsection{Noise}\label{sec:noise}
\subsection{Noise Distribution for Visibility}\label{sec:noise_dist_theory}

Noise in the cross-power spectrum is the difference of an observation and the ensemble average: 
$\tilde r_k - \langle \tilde r_k \rangle_{\rm n}$.
Typically, the observed spectrum is an average over a number of individually-formed spectra,
as describd by Equations\ \ref{eq:rtau_def} and \ref{eq:Fourier_transf_xpower}.
Consequently, the Central Limit theorem suggests that the noise follows a Gaussian distribution.
Because the cross-correlation function is complex, this distribution is an 
elliptical Gaussian distribution in the complex plane.

When the signal is completely absent, as when the pulsar is off,
one expects that $\tilde \rho_k=0$ and that $\tilde r_k$ consists of noise drawn from a zero-mean,
circular complex Gaussian distribution.
Our observations match this expectation closely,
as we discuss in Section\ \ref{sec:empty_dist}.

% \subsubsection{Noise Distribution}

If the signal is present, then the Dicke Equation describes the contribution of self-noise.
This equation states that 
the error $\delta T$ in measurements of antenna temperature varies with total system temperature $T$,
including the contribution of the source \citep{Dic46}:
\begin{equation}
\label{eq:dicke_eq}
(\delta T)^2 = {\frac{T^2}{ N_{\rm obs} }}.
\end{equation}
Here, $N_{\rm obs} = \Delta\nu\times\Delta t$ is the number of samples,
for an observed bandwidth $\Delta\nu$ and integration time $\Delta t$.
The analogous expression holds for interferometric visibility \citep{tms86}.
More generally, this equation describes the noise in the sample variance for draws from a Gaussian distribution. 

The real and imaginary parts of statistical averages of $\tilde r_k$ suffice to estimate both the signal and the noise \citep[][Equations 11, 18, 19]{gwi06}:
\begin{align}\label{eq:noise_alpha_rho}
\langle \tilde r_k\rangle_{\rm n} &= \tilde \rho_k  \\
\delta \tilde r_k \, \delta \tilde r_k^* \equiv \langle  \tilde r_k  \tilde r_k^* \rangle_{\rm n} -  \langle  \tilde r_k \rangle_{\rm n}\langle \tilde r_k^* \rangle_{\rm n} &= {\frac{1}{N_{\rm obs}}} \tilde \alpha_{Xk}  \tilde \alpha_{Yk}  \nonumber \\
\delta \tilde r_k \, \delta \tilde r_k \equiv \langle  \tilde r_k  \tilde r_k\rangle_{\rm n} -  \langle  \tilde r_k \rangle_{\rm n}\langle \tilde r_k\rangle_{\rm n} &= {\frac{1}{N_{\rm obs}}} \tilde \rho_k  \tilde \rho_k  \nonumber.
\end{align}
Again, $N_{\rm obs}$ is the number of samples gathered, per spectral channel. 
The subscripted angular brackets $\langle ...\rangle_{\rm n}$ indicate an average over many realizations of noise.
As the first expression indicates, $\tilde \rho_k$ is the cross-power spectrum, averaged over an ensemble of statistically-identical realizations of noise.
Analogously, $ \tilde \alpha_{\mathrm{X}k}$ is the autocorrelation function at station $X$, and $\tilde \alpha_{Yk}$ is the autocorrelation function at $Y$.
We assume single-sideband operation.  
Here we also apply the ``wrap'' assumption,
discussed in Sections\ \ref{sec:corrfunc_spectra} and\ \ref{sec:empty_gate_corr}.
The autocorrelation spectra are always real and are often identical between stations after calibration, with offsets for background noise.
For a scintillating source observed on a long baseline, the spectra need not be identical, because the stations may lie in different scintillation elements
in the observer plane.

To help visualize the distribution of noise in 
Equation \ref{eq:noise_alpha_rho}, we divide the noise into components parallel with and perpendicular to the
phase of the average visibility. The expression for noise then takes the form:
\begin{align}
\label{eq:noise_perp_para}
\tilde \rho &= |\tilde \rho| e^{i\phi}  \\
\sigma_{||}^2 &= {\frac{1}{2 N_{\rm obs}}}\left( |\tilde \alpha_{X}| |\tilde \alpha_{Y}| +  |\tilde\rho|^2 \right)\nonumber \\
\sigma_{\perp}^2 &= {\frac{1}{2 N_{\rm obs}}}\left( |\tilde \alpha_{X}| |\tilde \alpha_{Y}| -  |\tilde\rho|^2 \right)\nonumber. 
\end{align}
Here, $\phi$ is the phase of the average visibility.
We have omitted the subscript $k$ for clarity.
For most interferometric observations, 
the intensity of the source is constant over the observer plane: $\tilde \alpha_X = \tilde\alpha_Y$.
The difference between $\sigma_{||}$ and $\sigma_{\perp}$ then produces an elliptical distribution of noise,
with major axis aligned with the average phase $\phi$.
Usually, background noise contributes a constant offset to $\alpha$,
and the source contributes the rest.
This form is quite general: it holds for any interferometric observations of a noiselike source,
not just observations of a scintillating pulsar.

\subsection{Noise for a Scintillating Source}

\subsubsection{Distribution of Noise}

A strong, scintillating source provides a good laboratory for the study of self-noise
because it provides many independent observations of visibility,
with different flux densities and (if the baseline is long) phases,
under identical conditions.
For a scintillating, pointlike source,
with flux densities $I_X$ and $I_Y$ at stations $X$ and $Y$,
and background noise equivalent to flux densities $n_X$ and $n_Y$,
the signal and noise are \citep[][Equations 5-7]{gwi11b}:
\begin{align}
\tilde \rho &=  \sqrt{I_X I_Y } e^{i \phi_{\rm s}}   \label{eq:scintnoise} \\
\sigma_{||}^2 &= {\frac{1}{N_{\rm obs}}} \left\{ {\frac{n_X n_Y}{2}} + {\textstyle{\frac{1}{2}}} (n_Y I_X + n_X I_Y) + I_X I_Y\right\} \nonumber \\
\sigma_{\perp}^2 &= {\frac{1}{N_{\rm obs}}} \left\{ {\frac{n_X n_Y}{2}} + {\textstyle{\frac{1}{2}}} (n_Y I_X + n_X I_Y)\right\} \nonumber
\end{align}
The visibility phase, $\phi_{\rm s}$, arises from phase differences between the pair of scintillation elements in the observer plane.
The variance of measurements of $\tilde r$ at that phase is $\sigma_{||}^2$; the variance at quadrature is 
$\sigma_{\perp}^2$.
The subscript for channel $k$ is omitted; all quantities are for one spectral channel.
Note that in this expression, 
the autocorrelation spectra at the two antennas $\tilde \alpha_{Xk} ,  \tilde \alpha_{Yk}$ differ both because the 
intensity of the source at the two antennas may differ: $I_X \neq I_Y$,
and because of different noise at the two antennas: $n_X \neq n_Y$.

\subsubsection{Short-Baseline Limit}\label{sec:short_baseline_limit}

For a short baseline, both antennas will lie within the same scintillation element.  In this case, $\phi_{\rm s}\rightarrow 0$, and $I_A=I_B\equiv I$.
However, the variance of the noise at the two antennas may still be different.
We can then express the distribution of noise in the form \citep[][Equation 8]{gwi11b}:
\begin{align} \label{eq:dicke_complex}
\sigma_{||}^2 &= b_0 + b_1 I + b_2 I^2  \\
\sigma_{\perp}^2 &= b_0 + b_1 I  \nonumber \\
I &\equiv |\tilde \rho |  \nonumber
\end{align}
In this expression, $b_2=1/N_{\rm obs}$.
This expression holds even if the source is mildly resolved by the scattering disk, because the normalized,
ensemble-averaged visibility remains nearly 1 \citep{gwi01}.

\subsubsection{Long-Baseline Effects}\label{sec:long_baseline_effects}

If the baseline is long compared with the scale of the scintillation pattern in the observer plane, 
then the scintillations differ at the two stations. Thus, $I_X\neq I_Y$.  
Nonetheless, inspection shows that the constant term, $b_0$, and the quadratic term, $b_2$, in Equation\ \ref{eq:dicke_complex}
still describe the behavior of noise correctly. The linear term, $b_1$, does not.  However, it converges to the same form
in an average over many scintillation elements with the same visibility $\tilde\rho$, for a point source.
This is seen by extending the calculation of visibility in \citet{gwi01} to include the intensities at the two antennas,
most easily by numerical calculation.
However, this average over realizations of scintillations converges much more slowly than the average over realizations of the noise.

If the baseline is long and the source is resolved by the scattering disk,
then $| \tilde \rho |^2 < \tilde \alpha_{X}  \tilde \alpha_{Y}$,
so that $|\langle \tilde r\rangle_{\rm n}|^2< I_X I_Y$.
Consequently, $\sigma_{\perp}$ will acquire some of the quadratic behavior of $\sigma_{||}$ in Equation\ \ref{eq:scintnoise}.
For a small source size, the effect is second-order in the size parameter 
and will become apparent only when the distribution of visibility is already significantly distorted \citep{gwi01}.

\subsubsection{Effects of Variability, Quantization, and Correlation}

The flux densities of pulsars in general, and the Vela pulsar in particular, vary intrinsically from one pulse to the next and within pulses \citep{kri83,joh01,Kra02}.
Variations on timescales shorter than the time to accumulate one sample of the spectrum, the ``accumulation time,''
lead to correlations of noise between spectral channels and increase the source noise contribution to $\sigma_{||}^2$ \citep{gwi11a}. 
If we parametrize these variations by $\delta I/I$, with $\langle \delta I \rangle=0$,
then the quadratic coefficient, $b_2$, in Equation\ \ref{eq:dicke_complex} becomes, 
\begin{equation}\label{eq:b2_amp_variation}
b_2 = \left( {\frac{\delta I}{I}}\right)^2 + \frac{1}{N_{\rm obs}}.
\end{equation}
The other coefficients, $b_0$ and $b_1$, are unchanged.  
This calculation follows Section 3.3.2 of \citet{gwi11a},
but with $\beta-1=\delta I/I$ averaging to zero over the suite of observations rather than over the accumulation time for a single spectrum.

Digitization, or more precisely quantization during digitization, also affects the noise.
If the correlation is not extremely strong, $\rho < 0.5$, 
and if the data are viewed in the spectral domain,
then the effects of quantization can be represented as a change in gain and a spectrally-constant offset.
This offset is often termed ``digitization noise'' and contributes to the values of $b_0$ and $b_1$, as one would expect.
The expressions for noise take the same forms as in Equations\ \ref{eq:noise_alpha_rho} through \ref{eq:scintnoise} above,
but with corrections to station gains and noise levels
\citep[see Equations 56, 57 of][]{gwi06}.

Correlations among spectral channels also characterize noise.
For sources of constant intensity, noise is uncorrelated between spectral channels, under the ``wrap'' approximation \citep{gwi06}.
Relaxation of that assumption can lead to an observable correlation of noise between channels, as we discuss further in Section\ \ref{sec:empty_gate_corr}.
Variation of flux density within the accumulation time for a single spectrum can also lead to significant correlations of noise \citep{gwi11a}.

\section{OBSERVATIONS, CORRELATION, AND CALIBRATION}\label{sec:observations}

We observed the Vela pulsar on
10 Dec 1997 using a network
comprising antennas at Tidbinbilla (70-m diameter),
Mopra (22-m),
Hartebeesthoek (26-m),
and the VSOP spacecraft (8-m).
The observations began at 14:15\ UT and ended at 22:40\ UT,
for a time span of 8:25.
The observing wavelength was 18\ cm.
We observed left-circular polarized radiation.
We recorded two 16 MHz frequency bands (IFs) at each antenna,
both as upper sidebands.
The bands spanned 1634 to 1650\ MHz (IF1) and 1650 to 1666\ MHz (IF2).
The data were digitized (quantized and sampled) at recording time,
thereby characterizing the electric field with a sign bit and an amplitude bit.
The data are thus 4-level, or 2-bit, quantized.

During the observations, the interferometer baseline from Mopra to Tidbinbilla had a projected length of approximately 400\ km.
The baseline from Hartebeesthoek to Tidbinbilla had a projected length approximately 9{,}400\ km.
% 0.593759  to 0.663845 -> orbit 1
%  0.864881 to 0.884025 -> orbit 2
The baseline from the VSOP spacecraft to Tidbinbilla had length of approximately 27{,}000\ km during the first period of data, 
from 14:00:56 UT
to 15:55:56 UT,
when the spacecraft was near apogee;
and of approximately 22{,}000\ km during the second period of data, 
from 20:45:25 UT
to 21:13:00 UT,
near the following apogee.
The first period spanned a longer time period and showed more homogeneous statistics.

We correlated the data with the Canadian S2 VLB correlator \citep{car99}.
This correlator is a reduced-table 4-level correlator;
in other words, the lowest-level products are ignored \citep{hag73}.
We correlated each IF separately with 8192 lags to form
a cross-correlation function.

We correlated the signal from the pulsar in 6 gates, synchronized with the pulsar's period of approximately 89\ msec.
Each gate was 1\ msec wide.
The first 5 gates covered the pulse.
The sixth gate was located far from the pulse,
where the pulsar was ``off".
Because of interstellar dispersion, each gate covered a range of pulse phases.
%Propagation through the interstellar plasma disperses the pulse in frequency,
%so that a single gate actually covers a range of pulse phases.
Individual pulses also 
vary in intensity. We averaged each spectrum over a number of pulses,
which reduced, but did not completely eliminate, this variation.
We averaged the results of the correlation for 2\ sec, or approximately 22.4 pulsar periods,
except on the baselines to the spacecraft, which we averaged
for 0.5\ sec, or approximately 5.6 pulsar periods. 
The pulsar was strong enough to contribute to the system temperature at the antennas;
this contribution affected the noise through settings for the digitizers at the antennas, particularly at the most sensitive antenna.

For a reduced-table 4-level correlator,
the optimal level settings 
are $v_0=\pm 0.90$ standard deviations, with weighting $n=3$ \citep{coo70}.
Because the intensity of the pulsar varies greatly during the pulse,
and because the quantizer levels were adjusted every 10 sec
to optimal values for the previous 10-sec period,
the quantizer levels $\pm v_0$
were not at this optimal setting in each gate.
Table\ \ref{table_quantizer_v0} gives the levels 
in Gate 1
(where the most variation occurred) 
and in the empty gate (which provides the most contrast).
For Tidbinbilla, the largest and most sensitive antenna,
the levels changed dramatically and the standard deviation was much greater when the pulsar was ``on'' because 
the pulsar made a large, variable contribution to system temperature.
The variations are particularly large in Gate 1, at the leading edge of the pulse \citep[see][]{kri83}.
In contrast, for the VSOP spacecraft, the smallest antenna,
the levels were the same on and off pulse, because the contribution of the pulsar was insignificant;
the large standard deviation of $v_0$ arose from a trend over the time span of the observations. 
For the medium-sized antennas, Hartebeesthoek and Mopra,
the levels changed between ``on'' and ``off'' gates,
but the standard deviations remained approximately the same.
The correlator corrected for these changes in $v_0$ when estimating the 
cross-power spectrum $\tilde\rho$;
however, differences of $v_0$ among gates changes the properties of the noise in those gates \citep[see][]{gwi04,gwi06}.
We discuss these effect in Section\ \ref{sec:correlator_effects}.

The pulsar gates were so narrow that different lags accumulated different numbers of samples.
The $8192$\ calculated lags spanned a time
comparable to the width of a pulse gate: 
$(1 {\rm \ msec})\times (16\ {\rm MHz}) = 16000\ {\rm complex\ samples}$;
therefore, large lags accumulated fewer samples than small lags. 
The correlation function, $r_{\ell}$, was correctly normalized
by the number of samples contributing at each lag (Equation \ref{eq:rtau_def});
thus, the average of the correlation function $\langle r_{\ell}\rangle$ 
was the same as would have
been measured with uniform sampling.
However, the noise varied with lag.
% However, the noise in the measured $r_{\ell}$ was greater at large lag, because of the fewer samples.

We 
Fourier-transformed the cross-correlation functions to form cross-power spectra.
Because the data were recorded in single sidebands,
the cross-power spectra contained 8192 channels with signal,
each with bandwidth 1.95\ kHz.
The phase of the cross-power spectrum included instrumental effects, primarily observational and instrumental
delays and rates, varying slowly with time and frequency \citep{tms86};
and effects of scintillation, varying more rapidly, over the timescale and bandwidth of scintillation \citep{des92}.

The noise in a particular measurement of the correlation function was diluted by the integration time, but increased by the number of spectral channels.
We accumulated $16\times 10^6$ statistically-independent complex samples per second.
On Earth-based baselines, over a time of 2\ sec, we sampled 22.4 pulse periods, 
for an average net integration time of 22.4\ msec in one of our pulsar gates.  Our 8192-channel spectra then contained 44 independent complex samples per spectral channel for each integration period.
However, because of the truncation of the correlation function by the pulse gate
discussed in Sections\ \ref{sec:corrfunc_spectra} and\ \ref{sec:not_wrapping}, higher lags accumulated approximately only
half as many samples;
thus, the average number of samples per spectral channel was approximately 33.
On baselines to the VSOP spacecraft, the shorter integration time yielded 11 samples per channel and integration time 
for central lags, and approximately 8 samples per spectral channel.

Tones were injected into the signal, for calibration, at Hartebeesthoek.
These tones were separated by 511 channels (998\ kHz) and were 1 or 2 channels wide.
In cross-power spectra, they had the effect of increasing the noise in these spectral regions.
We removed the narrow spectral regions containing these spikes
from the spectra before continuing with analysis.

We removed the average delay and rate by fringe-fitting \citep[see][]{tms86}.
The fit included a fringe rate in time, a delay or slope of phase with frequency, and an overall phase offset.
For baselines between terrestrial antennas, 
we fringe-fit to (the central 7168 channels in frequency)$\times$(8 samples in time, or 16\ sec),
in the strongest gate, Gate 2.
We applied the fringe rate and delay to the other gates, but calculated the overall phase offset independently.
The results appeared to be nearly the same as fitting such a model to other gates that contained strong signal, such as Gate 1;
however, we prefer to use precisely the same model in weak and strong gates.
For this paper, the primary purpose of fringe-fitting was to remove all
instrumental effects, leaving only the effects of scintillation and those of
statistical noise in the data.

\subsection{Typical Data}\label{sec:typ_data}

\begin{figure}[t]
\epsscale{.80}
% fig 1:
% \plotone{/Users/carlgwinn/Desktop/VSOP_True/gates/proslide3.pdf}
% \plotone{figures/proslide3.pdf}
\includegraphics*[width=0.98\textwidth]{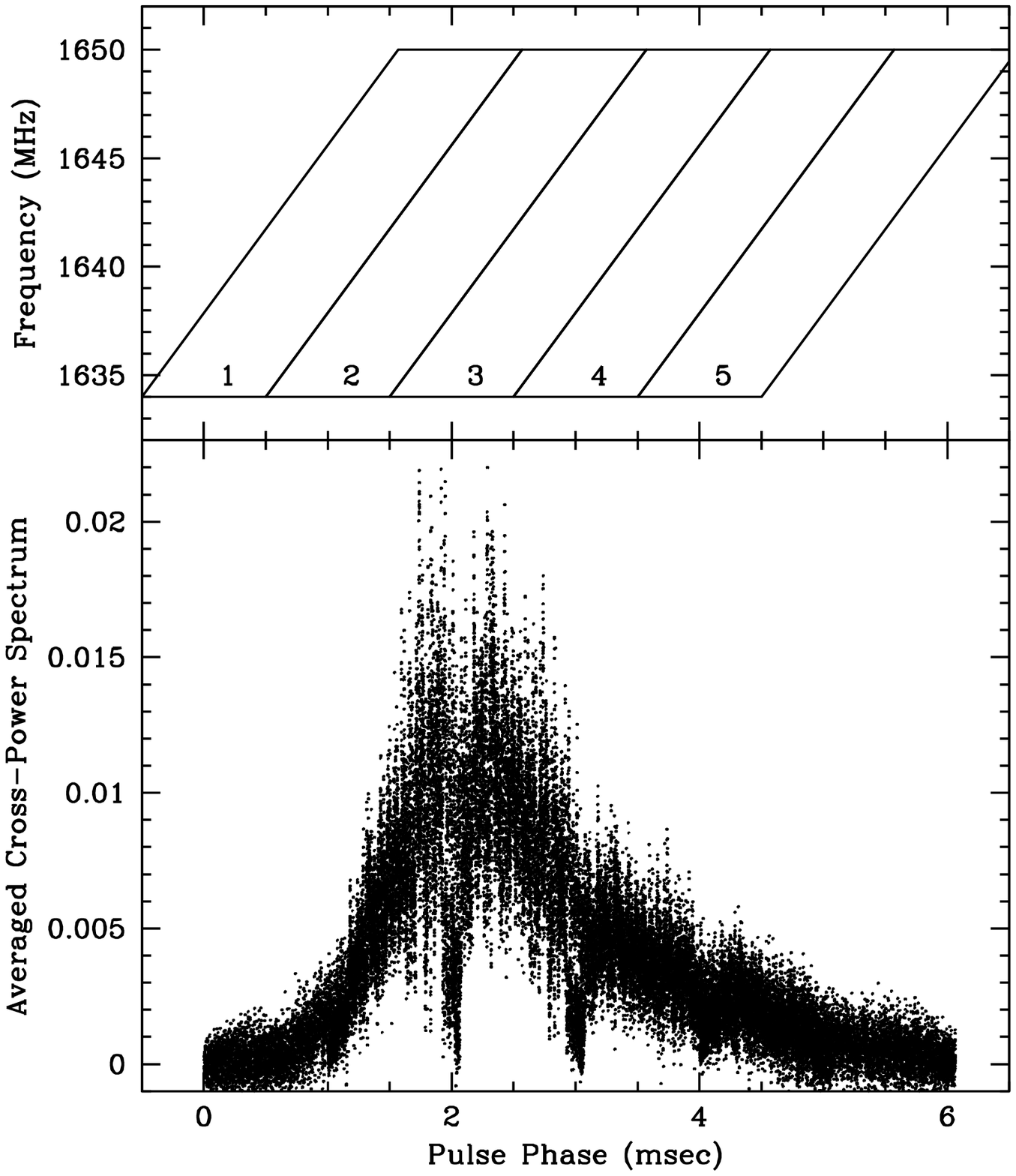}
\figcaption[]{
Gates across the pulse.
Lower panel: Averaged cross-power spectrum 
for the Mopra--Tidbinbilla baseline, in IF1,
displayed as a function of pulse phase.
Upper panel: Gates displayed with pulsar phase and frequency.
Gates are defined in frequency and time, and appear as parallelograms in this plot because of dispersion.
Each point in the lower plot shows one spectral channel,
at the center of 1 msec of pulse phase.
Spikes show scintillations during the 5-min observing period.
Dips in the pulse profile show rolloff of gain at the edges of the frequency band,
in each gate.
\label{proslide}}
\end{figure}

Figure\ \ref{proslide} shows some sample data, along with the scheme of
pulse gates.
To produce this figure,
we averaged the real part of a segment of data for IF1 on the short Mopra--Tidbinbilla baseline, 
from 14:15:05 UT to 14:20:00 UT.
This averaging reduced the depth of scintillation.  
We display the resulting average 
in the 5 gates as a function of pulse phase, including pulse gate and spectral dispersion.
Each gate sampled the IF bandwidth over a short range of pulse phase; however, 
because of pulse dispersion, the low-frequency end of the gate sampled
earlier parts of the pulse than the high-frequency end. 
The plot gives a rough idea of the pulse profile, although the effects of scintillation are still quite
large and each sample is averaged over the 1 msec gate.
As the figure shows, each phase in the pulse profile is represented twice,
at two different frequencies.  Each frequency is represented 5 times, in each of 
the gates (as well as in the empty 6th gate, outside the plot).

\begin{figure}[t]
\epsscale{.80}
% fig 2:
% chans 2560 to 2970 used for this figure
%% \plotone{/Users/carlgwinn/Desktop/VSOP_True/ahist_MT/drawnarr.pdf}
%\plotone{figures/drawnarr.pdf}
\includegraphics*[width=0.98\textwidth]{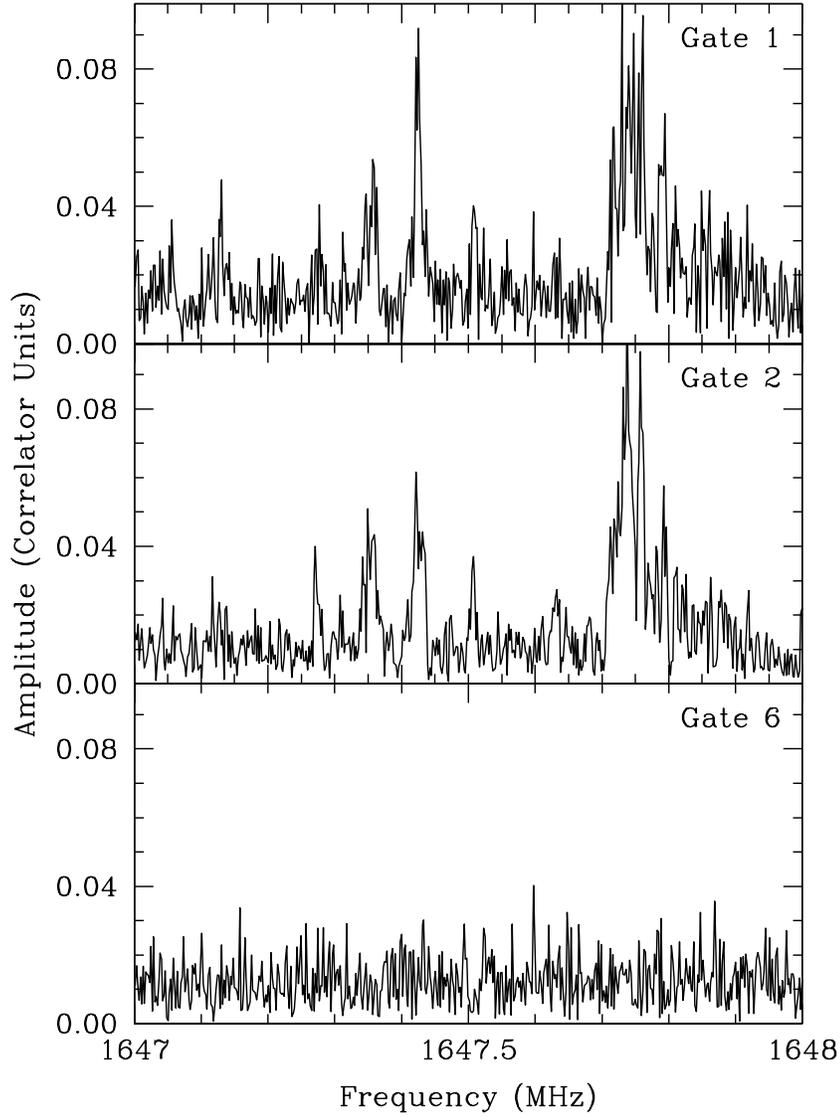}
\figcaption[]{
One frequency range of cross-power spectrum for the
Mopra-Tidbinbilla baseline in Gates 1, 2, and 6.
Spectra are averaged over 2 sec 
starting at 15:43:00 UT.
The first two gates
show nearly the same spectrum,
to within an overall factor,
because they sample the same time and frequency interval of the scintillation pattern.
Gate 6 contains only noise.
\label{typ_spect_narrow}}
\end{figure}

Figure\ \ref{typ_spect_narrow} shows cross-power scintillation spectra.
It compares the same 1-MHz spectral range for a single 2-sec integration, in 
Gates 1, 2, and 6 (off-pulse).
The scintillation appears as dramatic variations in amplitude,
with large amplitudes concentrated in a few spectral regions.
As measured from all of the Mopra-Tidbinbilla interferometric data,
the typical scintillation bandwidth was 15\ kHz (halfwidth at half-maximum of the autocorrelation function, in frequency)
\citep{gwi12}.
This frequency scale is apparent in the spectra.
Note, however, that noise modulates the scintillation
peaks on finer scales and introduces differences in detailed shapes of the peaks between gates.  
The typical scintillation timescale was 9\ sec ($1/e$ point of autocorrelation function, in time).
This timescale is longer than the integration time of the spectra. 
These scintillation scales are in good agreement with results extrapolated from single-dish measurements
by other observers at other frequencies \citep[see, for example,][]{Rob82,cwb85}.  
The spectrum in Gate 6 appears completely noiselike: indeed, the samples 
are drawn from a circular Gaussian distribution in the
complex plane, as discussed in Section\ \ref{sec:empty_dist} below.

\section{OBSERVATIONS OF THE NOISE DISTRIBUTION}\label{sec:noise_dist_obs}

\subsection{Strategy: Noise Estimates from Differences}\label{sec:strategy_binning_noise}

\subsubsection{Noise Estimates}

Differences between samples separated by much less than the 
scales of scintillation reflect noise and intrinsic variability; these differences can therefore be used to estimate properties of the noise. 
For example, consecutive samples in time can estimate both the signal (as their average) and the noise (as $\sqrt{2}$ times their difference).
More generally, we can estimate the signal and noise as averages over, 
and differences among, groups of nearby samples.
These estimates are unaffected by the scintillation if 
all of the samples are contained well within the characteristic frequency and time scales of the scintillation. 

Specifically, if we assume that the statistically-averaged signal $s_{k}$ 
is identical in $N$ samples $r_k(t_\ell)$, and that the noise $n_{k,\ell}$ is uncorrelated,
then we can estimate both the signal and the noise as,
\begin{align}
\label{eq:noise_differences}
s_{k} &= \frac{1}{N} \sum_{\ell} r_{k}(t_{\ell}) %\label{eq:multi_noise_estimate} 
\\
n_{k,\ell} &= \sqrt{\frac{N}{N-1}} (r_{k}(t_{\ell}) - s_{k} ) \nonumber,
\end{align}
where the index $\ell$ runs over the $N$ samples compared.
By binning the estimated noise, $n_{k,\ell}$, according to the estimated signal, $s_{k}$, 
and then estimating the variance in each bin, we can identify the changes of the distribution of noise as a function of signal,
and so identify the 3 coefficients $\{b_0, b_1, b_2\}$ in Equation\ \ref{eq:dicke_complex}.

\subsubsection{Effects of Scintillation, Variability, Correlation, and Binning on Noise Estimates}\label{sec:binning_effects}

The estimate of noise in Equation\ \ref{eq:noise_differences} does not include four effects that can contribute to noise: scintillation, amplitude variations, correlation of noise, and binning.
Scintillation changes the spectrum over the scintillation bandwidth and timescale,
so that differenced samples do not have identical averages.
This change will increase the estimated variance by an amount proportional to the square of the signal, 
and so will affect estimates of the coefficient of the quadratic term, $b_2$.
We limit the span of averages to less than the scintillation time to minimize this effect.

Scintillation also introduces differences between the autocorrelation functions at the two antennas,
as discussed in Section \ref{sec:long_baseline_effects} above.
The difference grows as the square of baseline length, divided by the scale of the diffraction pattern for short baselines.
As discussed in Section \ref{sec:long_baseline_effects},  an average over many scintillation elements recovers the behavior
given by Equation\ \ref{eq:dicke_complex}.
% Thus, binning the noise estimates with this average yields the same behavior, and can recover estimates of $\{b_0, b_1, b_2\}$.
However, this average over scintillation elements
converges much more slowly than the average over realizations of the electric field used for short baselines.

Amplitude variations of
the Vela pulsar are significant \citep{kri83,joh01,Kra02}.
Amplitude variations of the pulsar produce differences between spectra and thus mimic effects of noise.
If the amplitude can be estimated from individual spectra, the effects can be removed \citep{gwi11b}.
In the observations reported here, rapid variations in time appear as spectral variations because of dispersion,
and the signal-to-noise ratio within a narrow region of the spectrum is too low to estimate amplitude reliably.
However, the intrinsic variability of the pulsar is reduced by our averaging over 22 or 23 pulses.
% Consequently,
% we expect intrinsic variability to have little effect except near the beginning and end of the pulse, where intrinsic modulation is greatest \citep{kri83}.
% Comparisons with variability estimated from the distribution confirm this \citep{gwi12}.

Correlation of noise in nearby spectral channels can arise from pulse gating, as 
discussed in Section\ \ref{sec:empty_gate_corr} below.
Larger correlations arise from variations of intensity on timescales shorter than the time to accumulate a single sample
of the spectrum \citep{gwi11a,gwi11b}.
For these observations, this timescale is $8192/16\ {\rm MHz} =0.512\ {\rm msec}$.
The Vela pulsar shows significant variations on shorter timescales \citep{kri83,joh01,Kra02},
so we expect these correlations to be significant.
Therefore, we do not use differences among spectral channels to estimate noise, but only differences among times.
(An exception is the VSOP-Tidbinbilla baseline discussed in Section\ \ref{sec:noise_dist_obs_long}, where 
low signal-to-noise ratio demands differencing in frequency as well as time.)

Binning of the noise estimates by signal requires an accurate estimate of signal $s_k$.
Even if the noise estimate is correct, 
it may be assigned to the wrong bin, altering the form of the estimated distribution.
This problem is most severe when the distribution of signal varies within the span of the error in $s_k$.
In this case, the noise can ``leak'' into the wrong bin. 
Consequently, low signal-to-noise ratio drives the analysis to few, large bins.
The problem can be lessened by integrating over a longer time interval, so that the average is better determined. 
These intervals must be smaller than the scales of variations from scintillation, which can mimic noise.

Alternative analyses, involving a model for the distribution of flux density of the scintillating source, 
avoid use of bins completely and so eliminate ``noise leakage.''
Such analyses can estimate all parameters, including noise parameters, simultaneously \citep[as in][]{gwi12,Joh12b}.
These methods are also insensitive to correlations between spectral channels. 
However, the differencing and binning analysis presented here allows straightforward 
visualization of the noise distribution,
as presented in Sections\ \ref{sec:noise_ellipse} through \ref{sec:noise_dist_obs_long} below,
without any assumption about the underlying distributions of visibility or noise.
% This analysis provides means of exploring our noise model for various baselines without a great many assumptions.

\subsection{Observations: Noise in the Complex Plane}\label{sec:noise_ellipse}

\begin{figure}[t]
\epsscale{.80}
% fig 3:
%\plotone{figures/ellipses_HT_ch2g2_2.pdf}
\includegraphics*[width=0.98\textwidth]{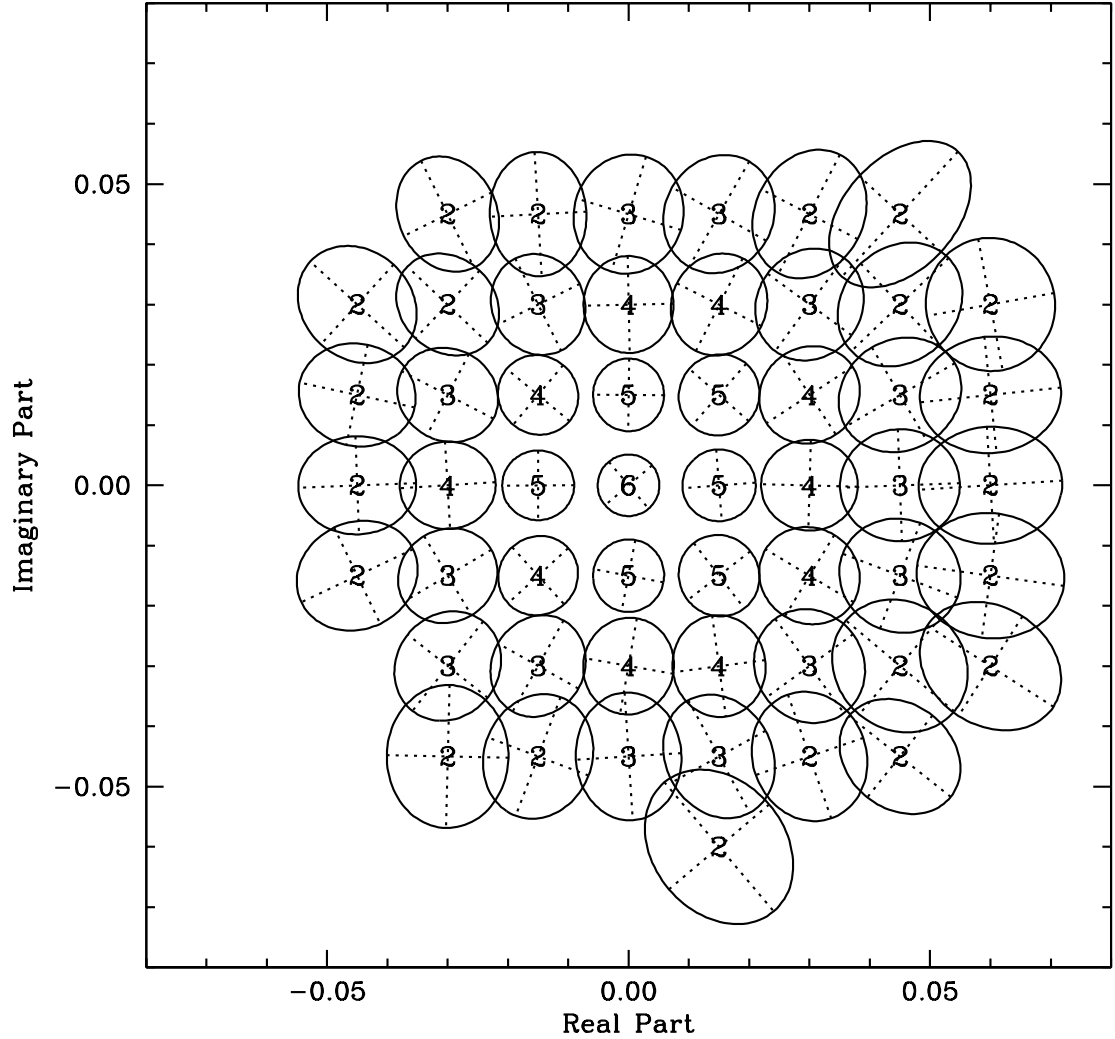}
\figcaption[]{
Distributions of noise, in bins of signal in the cross-power spectrum,
in IF 2 Gate 2 Channels 2048 to 3072 on the Hartebeesthoek--Tidbinbilla baseline. 
For each bin in real and imaginary part
of average signal, the ellipse shows one-half the 
standard deviations of noise within that bin. 
Ellipses are plotted for bins that contain $N>100$ noise values. 
Labels are $\log_{10} N$.
\label{adiff_ellipses}}
\end{figure}

The interferometric visibility is complex, and the distribution of noise varies with it over the complex plane.
Figure\ \ref{adiff_ellipses} shows estimated noise 
as a function of average visibility in the complex plane,
for the Harte\-beesthoek--Tidbinbilla baseline.
This baseline was long enough to span much of one scintillation element,
so that the interferometric phase varies over $2\pi$.
Moreover, the antennas are large, so the resulting moderate signal-to-noise ratio allows the complex plane to be divided into many bins.
Note that although the distribution extends over $2\pi$ in phase, it is concentrated toward the right:
the average visibility lies on the positive real axis.

We used data 
from IF2, Gate 2, channels 2048-3072,
for the entire time span of the observation on that baseline.
We formed spectra, and fringed the data, as described in
Section \ref{sec:observations} above, to align the average visibility with the positive real axis.
We then differenced pairs of consecutive time samples to find the 
estimated noise $n_{k}$ and signal $s_{k}$.
The data were binned by real and imaginary parts of the average signal $s_k$,
with bin increments of 0.015 correlator units.
For bins containing more than 100 samples, we found error ellipses 
from the standard deviation of our estimates of the complex noise $n_{k}$ in that bin.
% The data are complex, so three parameters describe the two principal axes and covariance in the two dimensions.
The figure shows these ellipses.
The displayed ellipses extend to one-half standard deviation of the noise
in each bin, to reduce confusing overlap.

The error ellipses have the form that Equation\ \ref{eq:dicke_complex} suggests,
with size increasing with distance from the origin,
indicating increasing noise with increasing signal amplitude.
Ellipses close to the origin are nearly circular, as they must be:
the noise is independent of signal phase, when
signal amplitude is near zero.
Both dimensions of the error ellipses grow with increasing signal;
however, the noise in phase with the signal grows faster, so that the ellipses
become elongated further from the origin.
% Major axes point toward the origin (along the phase of the signal), as expected.
Of course, amplitude variations in time, between pulses, would produce the same effect.
However, the contribution of such amplitude variations is small for our 2-sec averages.
%Some bins near the edges of the distribution have major axes that do not point directly
%toward the origin; 
%these are in regions where the population of points changes rapidly,
%and probably reflect noise near one edge of the bin, or 
%``spill over'' from adjacent bins, from imperfect separation of signal and noise.

\subsection{Observations: Noise on a Short Baseline}\label{sec:noise_dist_obs_short}

\begin{figure}[t]
\epsscale{.80}
% fig 4:
% \plotone{figures/fitem_ch2g1_5120_6144.pdf} -- see ahist_MT/binned/fitem/fitem_ch2g1_50120_6144.pdf
\includegraphics*[width=0.98\textwidth]{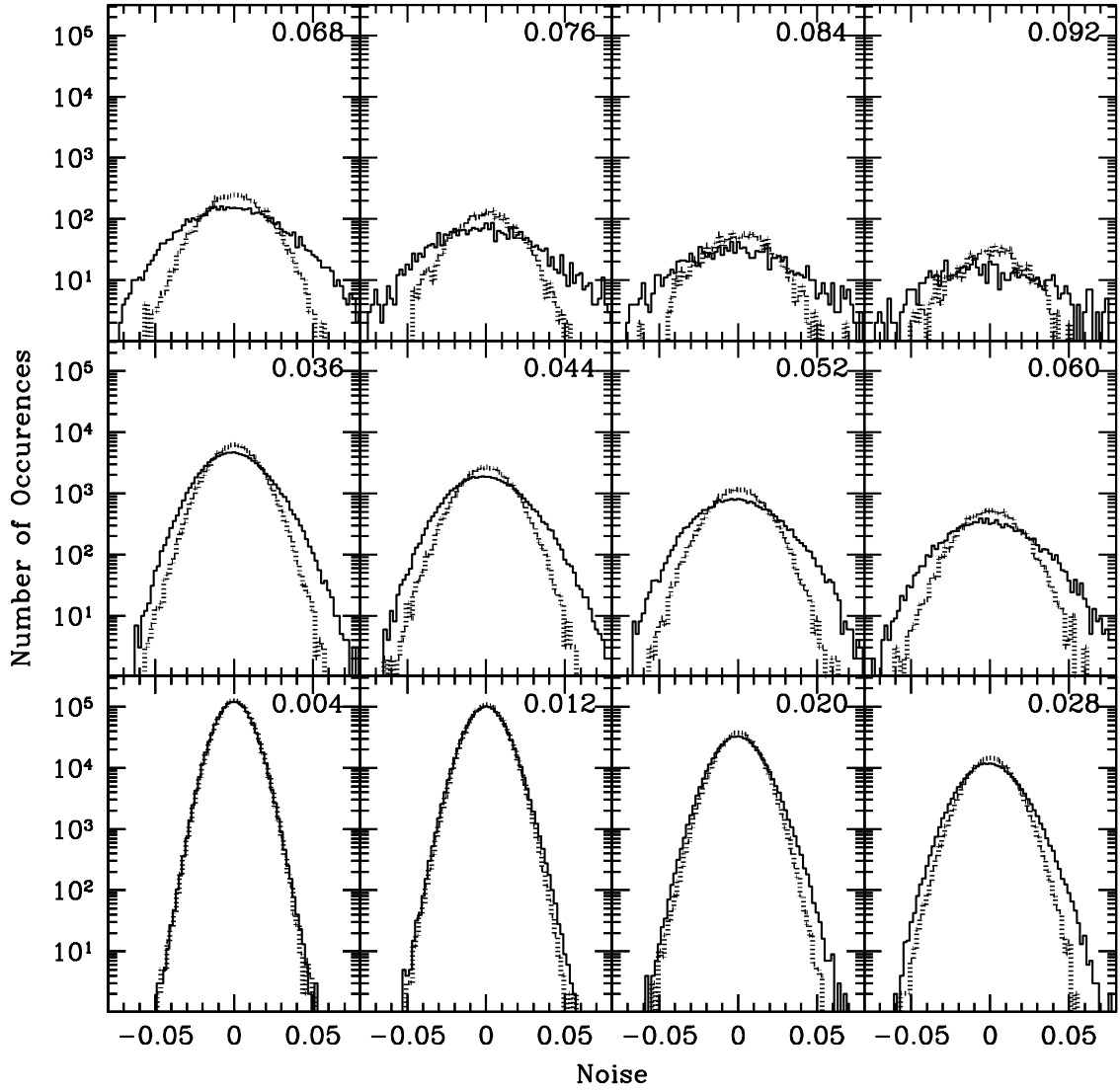}
\figcaption[]{
Distribution of noise in phase with the signal (solid) and in quadrature (dotted),
for bins in amplitude of the signal,
in IF 2 Gate 1 Channels 5120 to 6144 on the Mopra--Tidbinbilla baseline.
Noise was found by finding differences from the averages of 4 samples in time.
Numbers at top right of each panel show the  amplitude of the signal 
at the center of the bin.
Units are correlator units.
\label{adiff_binned_hists}}
\end{figure}

Figure\ \ref{adiff_ellipses} and the form of Equation\ \ref{eq:dicke_complex} 
suggest that we find noise parallel with and perpendicular to the signal.
Thus, as long as we consider perpendicular and parallel components separately,
we can group together noise estimates with the same magnitude of visibility.
Figure\ \ref{adiff_binned_hists} shows this analysis
for IF2, Gate 1 Channels 5120-6144 on the Mopra--Tidbinbilla baseline.
This short baseline provided a long observation, yielding
many individual measurements of noise.
The chosen gate and channel range are near the pulse peak,
where the source was strong and signal-to-noise ratio is high,
providing for many, relatively narrow bins.
As in Equation\ \ref{eq:noise_differences},
we averaged the signal over 4 samples, or 8 sec.
We then found noise by differencing individual samples from the average.
For each bin in signal amplitude, we present histograms of
noise in phase with the signal and at quadrature.

As the figure shows, noise was equal in the two directions at zero signal, 
at lower left,
and increased with signal amplitude to the right and upward.
Moreover,
noise increased more in phase with the signal, than in quadrature with it,
so that the widths of the two distributions in each panel diverge as signal increases.
The number of points decreased with increasing signal as well, representing the effect of the underlying distribution of visibility.
The vertical scales are logarithmic, so that a parabolic shape indicates a Gaussian distribution.
We do not display fits of Gaussian distributions to the histograms in
the figure because 
these are nearly indistinguishable
from the histograms.

Figure\ \ref{adiff_sigs} shows the variances of the best-fitting Gaussian distributions
to noise, for the data shown in Figure\ \ref{adiff_binned_hists}.
The noise in phase with the signal (shown by circles in the figure)
increases quadratically with signal amplitude,
whereas noise in quadrature with the signal (shown by crosses)
increases linearly.
This is precisely the behavior expected, as discussed in Section\ \ref{sec:noise_dist_theory} above.
We fit these two curves
with polynomials of the form given by Equation\ \ref{eq:dicke_complex}.
We assume that the coefficients (except the quadratic coefficient $b_2$)
are the same for $\sigma_{||}^2$ and $\sigma_{\perp}^2$.
The fit is to points in the range $0.006<I<0.056$.
Points at larger $I$ are based on a small number of samples.
In the lowest bin, noise is slightly higher than expected in both components,
compared with extrapolation from larger bins and the fit.
The noise increases in all directions away from this bin,
so that every point ``leaked'' from an adjacent bin tends to increase the noise,
as discussed in Section\ \ref{sec:binning_effects} above.
% The fit is excellent.
The y-intercept of the fit is close to the noise level estimated for the empty
spectral range of this gate, discussed in Section\ \ref{sec:empty_dist} below. 
The figure also shows, as a dotted line, results for noise in a fit to the global distribution of visibility discussed in \citet{gwi12}.
% Agreement of the two techniques is excellent.  
The two techniques usually agree well, for gates and spectral ranges with high amplitude,
so that leakage of noise into adjacent bins is small, and the distribution can be characterized well.

As a measure of the quality of these fits,
we note that 
in the absence of self-noise, all 16 points  
in Figure\ \ref{adiff_binned_hists} would have a single value $b_0$.
The mean square residual about the best-fitting single value is $6.5\times 10^{-1}$.
Our model included two additional parameters, $b_1$ and $b_2$, and reduced the mean square
residual to $2.4\times 10^{-3}$, or by a factor of $268$.
The F-ratio test gives the likelihood of this improvement occurring by chance of less than $10^{-12}$ \citep{Bev}.
Thus, the 3-parameter model is excellent.
If we adopt the parameters from \citet{gwi12}
shown by the dashed lines in Figure\ \ref{adiff_binned_hists},
the mean square residual is $3.3\times 10^{-3}$.
%The relatively larger residual may reflect the limitations of binning the data, as discussed in 
%Section\ \ref{sec:binning_effects} above.

The difference between  $\sigma_{||}$
and $\sigma_{\perp}$ 
reflects self-noise and effects of amplitude variations. 
The contribution of self-noise was $1/N_{\rm obs} = 1/33$, as discussed in Section\ \ref{sec:observations}.
As Equation\ \ref{eq:b2_amp_variation} shows, amplitude variations on
timescales longer than the accumulation time, approximately 1\ msec,
and shorter than the 2-sec integration time,
contributed to $b_2$.
For the data in the figure, 
we found $b_2=0.12$.
This value indicates that $(\delta I/I )^2= 0.09$, as would be expected after integrating over 22 or 23 pulses \citep{kri83,joh01,Kra02}.

\subsection{Observations: Noise on an Intermediate Baseline}\label{sec:noise_dist_obs_intermed}

\begin{figure}[t]
\epsscale{.80}
% fig 5: 
%\plotone{figures/adiff_7_c3_ch1g1_5120_6144_thick.pdf}
% MT_adiff_7_c3_ch1g1_5120_6144_thick.ps  see ahist_MT/binned/compare_noise/ch1g1_5/
% MT_adiff_7_c3_ch1g1_5120_6144_thick.pdf
\includegraphics*[width=0.98\textwidth]{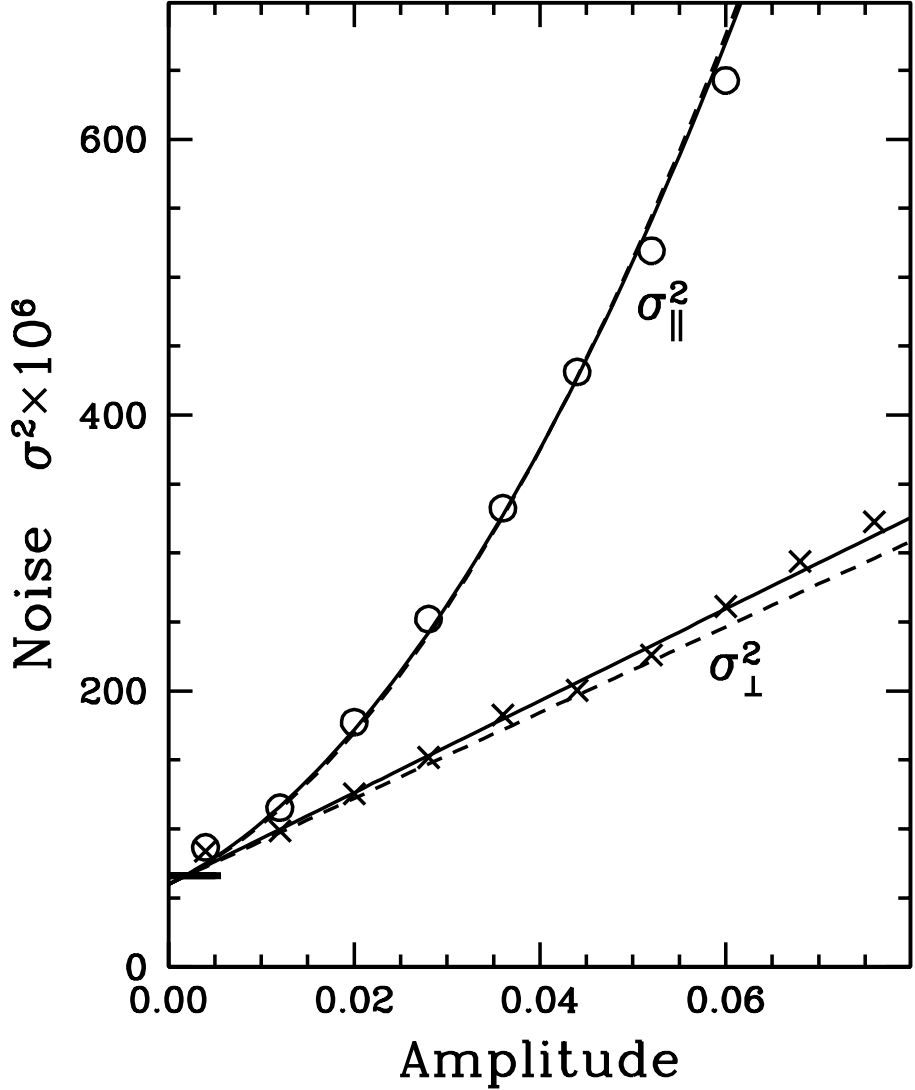}
\figcaption[]{
Variances of noise in phase with the signal (circles) and at quadrature (crosses)
estimated from differences of channels
from the average of 4 consecutive samples in time.
Data are from
IF 2 Gate 1 Channels 5120 to 6144 on the Mopra--Tidbinbilla baseline.
Solid curves show best-fitting line and parabola to the points,
with the fit demanding the same linear terms for both.
Leftmost and rightmost points were excluded from the fit.
The dashed curve shows the curve from a fit to the full distribution \citep{gwi12}.
The heavy tick 
on the y-axis shows noise 
in the spectral region of Gate 1 empty of signal (see Figure\ \ref{empty_dist}).
Units are correlator units.
\label{adiff_sigs}}
\end{figure}

\begin{figure}[t]
\epsscale{.80}
% fig 6: 
%\plotone{figures/fitfits_ch2g1_5120_6144.pdf}
% ./ahist_HT/binned/fitem/HT_fitfits_ch2g1_5120_6144_thick.pdf
\includegraphics*[width=0.98\textwidth]{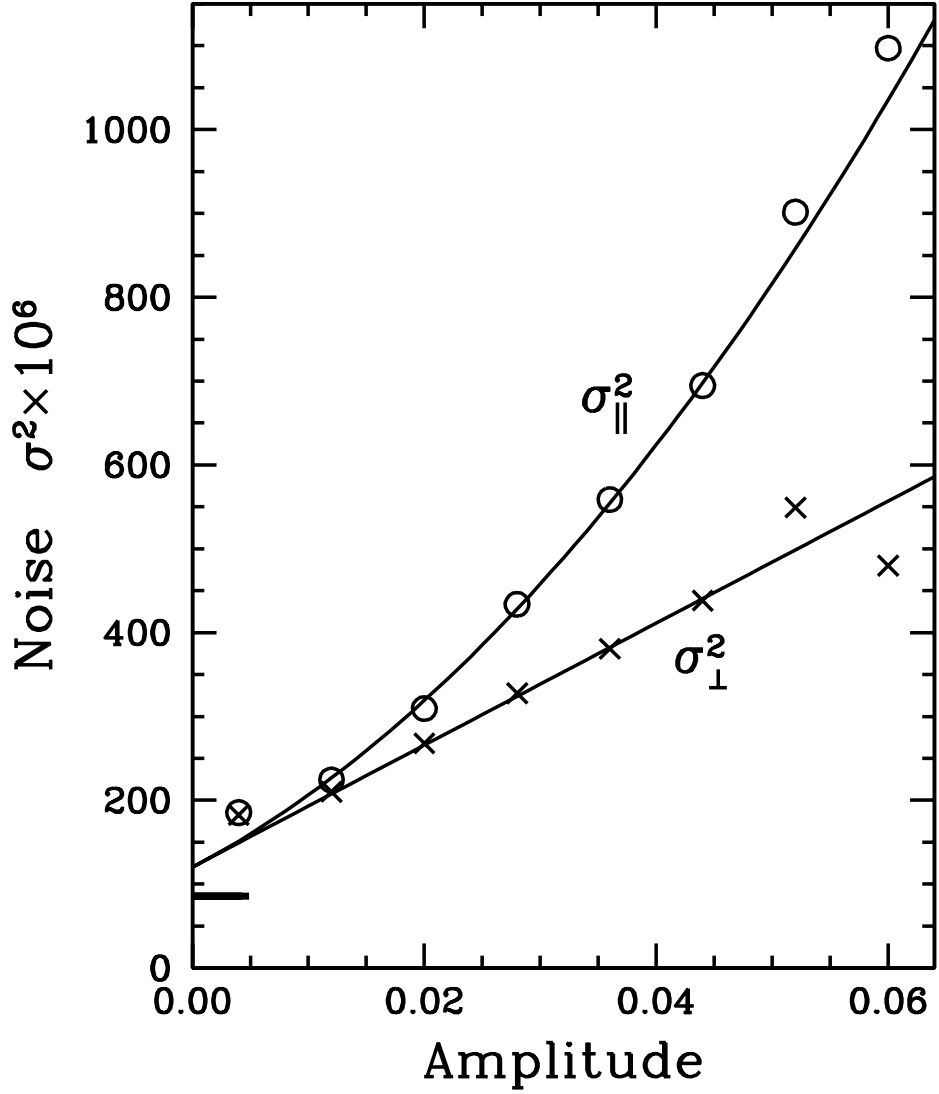}
\figcaption[]{
Same as Figure\ \ref{adiff_sigs}, but for the Hartebeesthoek-Tidbinbilla baseline,
for IF 2 Gate 1 Channels 5120 to 6144.
Again, the heavy tick 
on the y-axis shows noise 
in the spectral region of Gate 1 empty of signal.
\label{adiff_sigs_HT}}
\end{figure}

Figure\ \ref{adiff_sigs_HT} shows the behavior of noise on the Hartebeesthoek-Tidbinbilla baseline.
This baseline had projected length of over 9,000\ km during the observation;
however, it is intermediate in length in the sense that 
it is comparable to the length scale of the scintillation pattern.
% as we will discuss elsewhere (Gwinn et al. in preparation).
Moreover, the baseline was short compared with the baseline to the VSOP spacecraft.
We included data from the entire time span recorded, in IF2, Gate 1, channels 5120-6144. 
This is the same gate and frequency range as shown for the Mopra-Tidbinbilla baseline in Figure\ \ref{adiff_sigs}.
This range was chosen for comparison, and because Gate 1 allows comparison with an empty region of the gate.
We fit these two curves
with polynomials of the form given by Equation\ \ref{eq:dicke_complex}, to points in the range $0.006<I<0.056$.
Our 3-parameter noise model reduced the mean square residual over the 10 points in the fit,
relative to a single variance,
by a factor of 1112. The F-ratio test gives probability of this occurring by chance of less than $5\times 10^{-12}$.

Comparison of Figures\ \ref{adiff_sigs} and \ref{adiff_sigs_HT} shows interesting consequences of the longer baseline length.
The maximum amplitude was smaller, because as the baseline length approaches the scale of the scintillation,
occurrences of high intensity at both stations become less likely.
The contribution of background noise $b_0$ was smaller
and the self-noise was greater,
perhaps because Hartebeesthoek has larger area than Mopra (26\  m rather than 22\ m diameter).

%The point at smallest amplitude shows considerably higher noise than extrapolated from larger amplitudes.
%Simulations suggest that this results from `leakage'' of noise from other bins into this bin.
%This leakage is entirely from the next larger amplitude bin, which contains nearly as many samples.
%For the other bins, leakage is present in both directions, and cancels to some degree,
%although it can increase the $y$-offset of both curves, so that they do not interpolate to the value found in the empty portion of the gate,
%indicated by the heavy horizontal line on the $y$-axis.
The points at high amplitude show scatter about the fitted curves; 
for this intermediate baseline, 
the average for $b_1$ converges 
over some timescale intermediate between 
the many samples of electric field for a short baseline, and 
the many scintillations required for a long baseline.
%This slow convergence, and the relatively small number of samples at large visibility, contributes to the variation of the points about the expected form.
The first bin is elevated relative to the extrapolated, fitted curves,
and the $y$-intercept of the fitted curves lies above the noise found in the empty part of the gate.
This may result from leakage of noise adjacent bins,
and the different convergence statistics.

\subsection{Observations: Noise on a Long Baseline}\label{sec:noise_dist_obs_long}

As an example of noise 
on a long baseline, we analyze the long baseline from the VSOP spacecraft to
Tidbinbilla, using procedures similar to those for the short Mopra--Tidbinbilla baseline.
On this long baseline the phase of the cross-power spectrum
varied through many turns because of scintillation. % \citep{des92,gwi98}.
Instrumental phase variations were larger and more rapid than on the Mopra--Tidbinbilla baseline discussed above,
and of course the 8-m spacecraft antenna is smaller than Mopra.
Fringe-fitting the data, as described in Section \ref{sec:observations}, was challenging
because of the rapid variation of phase and rate from spacecraft motion, the large variations of phase with scintillation and the low average visibility.

For analysis of noise, we used fewer, larger bins in visibility so that the noise in a bin does not greatly exceed bin width.
For our tests, we used data from the first orbit, from IF1.
We used Gate 2, which is near the peak of the pulse and has high flux density across the observing band, to maximize signal and so ease fringe-fitting.
To improve the quality of the noise measurement, and to reduce leakage into adjacent bins,
we found averages for the real and imaginary parts of the visibility for spans of 4 channels in frequency, and 8 samples in time ($7.8\ \mathrm{kHz}\times 4\ \mathrm{sec}$).
The frequency span lies well within the scintillation bandwidth of 15\ kHz,
and the time span within the scintillation timescale of 9\ sec and the fringing time of 16\ sec.

\begin{figure}[t]
\epsscale{.80}
% fig 7:
%
\includegraphics*[width=0.98\textwidth]{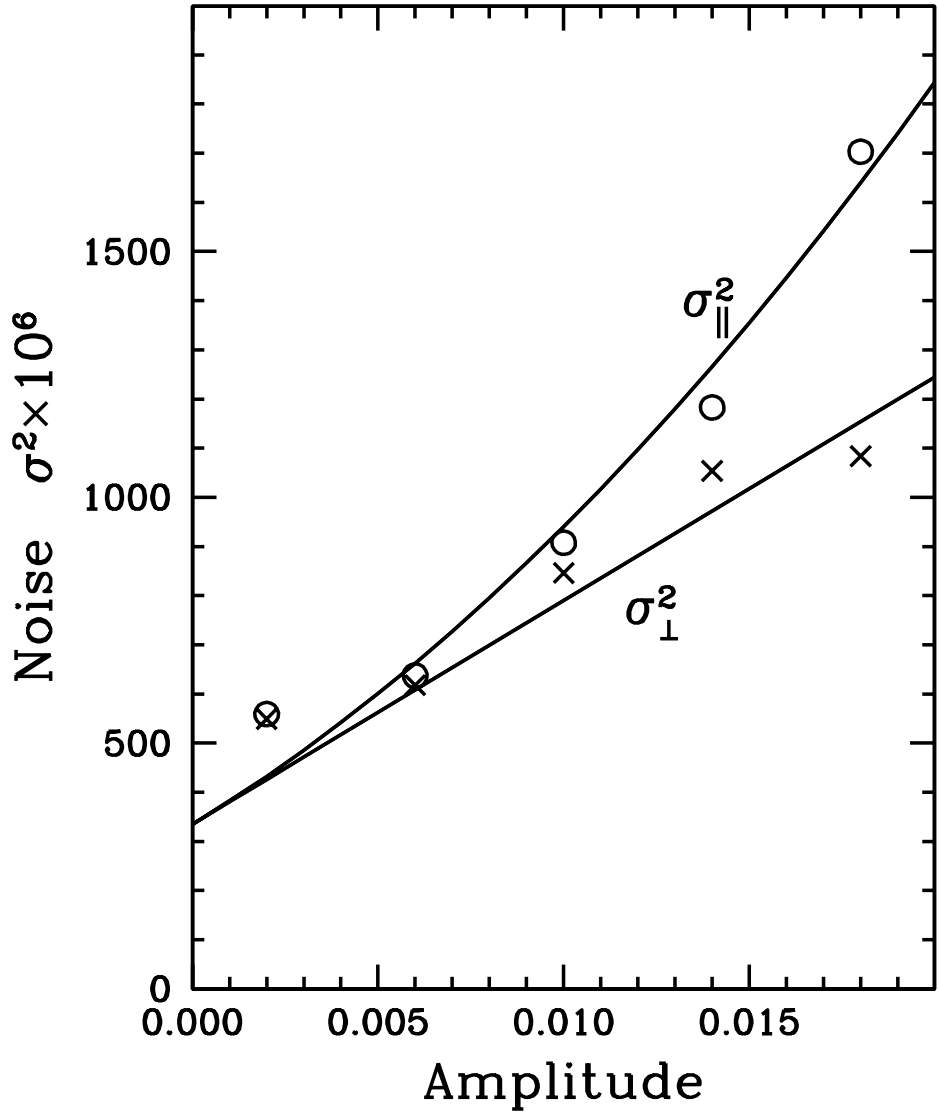}
\figcaption[]{
%xxx hfit_dat_ch1qg2_13_orbit1.pdf xxxxx
%xxxx ch1 gate 2 channels 1024 to 3079 , good timerange % $(ou[krbread].sd.ut>0.619634)&&(ou[krbread].sd.ut>0.638762)$
%xxx
Same as Figure\ \ref{adiff_sigs}, but the VSOP--Tidbinbilla baseline,
and in IF1Gate 2 in channels 1024 to 3079,  for Orbit 1 near apogee,
for 14:52:16 to 15:19:49 UT. 
Averaged signals from (4 channels in frequency) $\times$ (8 samples in time).
% No region empty of pulsar emission is available in Gate 2 for comparison with the $y$-intercept.
\label{adiff_sigs_VT}}
\end{figure}

The noise is different in phase and in quadrature with the signal.
Figure\ \ref{adiff_sigs_VT} shows the variances $\sigma_{||}$ and $\sigma_{\perp}$
plotted with the averaged magnitude of the visibility $|s_k|$.
We fit to the range $0.005 < |s_k| < 0.020$.
Again, quadratic and linear models, with linear terms identical, 
fit the data well for small signal amplitude.
The 3-parameter model reduces the mean square residual by a factor of 31,
relative to a single-parameter model.
The F-test gives probability of chance occurrence of less than $6\times 10^{-5}$.

The quadratic term is small compared with the linear term, in comparison
with the other baselines, 
as expected for the lower signal-to-noise ratio 
for this less-sensitive baseline.
The point at smallest amplitude shows considerably higher noise than that extrapolated from larger amplitudes,
again most likely from `leakage'' of noise from other bins into this bin.
As discussed in Section \ref{sec:long_baseline_effects},
the noise converges to the expected form only in an average over many scintillation elements.
This slow convergence, and the relatively small number of samples at large visibility, contribute to the variation of the points about the expected form.

\section{ANALYSIS: CORRELATOR EFFECTS}\label{sec:correlator_effects}
\label{sec::Correlator_Effects}

\subsection{Statistics of Noise in an Empty Gate}\label{sec:empty_dist}

\subsubsection{Variance: Short Baseline}\label{sec:empty_gate_dist_variance}

\begin{figure}[t]
\epsscale{.80}
% fig 8:
% \plotone{/Users/carlgwinn/Desktop/VSOP_True/ahist_MT/makehist/new_noise61/noise61_redo_more.pdf}
%\plotone{figures/new_noise61_redo.pdf}
\includegraphics*[width=0.98\textwidth]{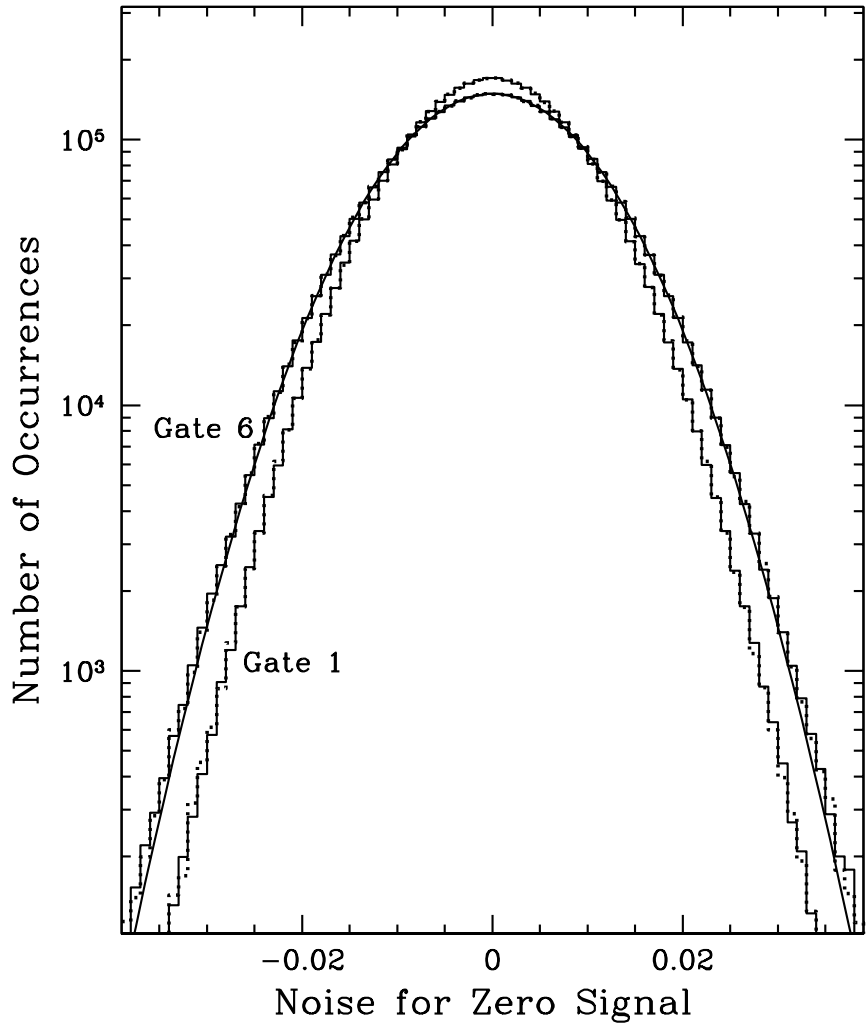}
\figcaption[]{
Distribution of noise with zero signal 
for the Mopra--Tidbinbilla baseline in Gate 6 Gate 1 
in IF1, channels 1024 to 2048.
Solid lines show real parts of cross-power $\tilde r_k$; dotted lines show imaginary parts.
Gate 6 is empty: the pulsar is ``off''.  Gate 1 contains 
pulsar emission at higher frequencies, but is nearly empty in
this spectral range.
Solid curve shows fit for a Gaussian distribution to the histogram, for Gate 6.
\label{empty_dist}}
\end{figure}

In spectral regions empty of signal,
the average value of the cross-power spectrum is zero,
and the noise closely approximates
a Gaussian distribution.
Figure\ \ref{empty_dist} shows two examples for the Mopra--Tidbinbilla baseline:
the distribution of 
the real and imaginary part of $\tilde r_k$
for IF1 channels 1024 to 2048 during the period 19:08:17\ UT to 21:12:52\ UT,
in Gates 1 and 6.
Gate 6 was correlated only for IF1.
The statistics of the real and imaginary parts are nearly identical, as expected for the noise,
as the figure shows.
The noise is clearly smaller in Gate 1, when the pulsar is ``on'' in another part of the gate.
The variances are $58.3\times 10^{-6}\ {\rm correlator\ units}^2$ for Gate 1, and $96.5\times 10^{-6}\ {\rm correlator\ units}^2$ for Gate 6.

The change in noise is an artifact of digitization \citep{gwi06}.
The reduction is the consequence both of the change in autocorrelation spectrum $\tilde \alpha$
and of the different levels of the quantizer
relative to the standard deviation of the signal (Table\ \ref{table_quantizer_v0}).
Equation\ 56 of \citet{gwi06} gives the noise in the absence of signal:
\begin{align}
b_0 &=\Gamma_C \frac{2 N}{N_{\mathrm{obs}}} \left(A_{X2} + B_X(\tilde\alpha_k-1)\right)\left( A_{Y2}+ B_Y(\tilde\alpha_k-1)\right),
\label{eq:quantizednoise}
\end{align}
This expression relates the notation of this paper on the left side of the equation, with that of \citet{gwi06} on the right.
Evaluation requires the autocorrelation spectrum $\tilde\alpha_k$, subject to the normalization condition: $\sum \tilde\alpha_k = 1$, where the sum runs over the $2N$ spectral channels \citep[][Equation 1]{gwi06}.
Effects of the change in the threshold for the quantizer $v_0$ are contained in the constants $A_{X2}$, $B_{X}$
for station $X$ and $A_{Y2}$, $B_{Y}$ for $Y$.
These quantities can be calculated from the statistics of the reduced-table 2-bit correlator with $n=3$ and the values for $v_0$ given in Table  \ref{table_quantizer_v0}.
The correlator-dependent gain $\Gamma_C$ parametrizes instrumental effects.
The number of spectral channels is $2 N$ and the number of observations is $N_{\mathrm{obs}}$.
For our observations, on a short baseline, $\frac{2 N}{N_{\mathrm{obs}}} \approx 33$ as discussed in Section \ref{sec:observations}.

For Gate 6, empty of any pulsar flux, the autocorrelation function was flat, and $\tilde\alpha_k \equiv 1$.
When the pulsar turns on in part of the spectrum,
the autocorrelation function in the off-pulse portion must fall,
because $\tilde\alpha_k$ is normalized.
%Under the plausible assumption that the level is constant, in voltage, 
%$\tilde\alpha_k \propto v_0^2$ in a channel $k$ of Gate 1 where pulsar flux density is zero, even if the pulsar is ``on'' in other channels.
This accounts qualitatively 
for the reduced noise level in the empty portion of Gate 1 relative to the completely-empty Gate 6,
although changes in $v_0$ also play a role.

A quantitative calculation of the autocorrelation spectrum when the pulsar is ``on'' is necessarily indirect.
We estimated the relative background noise of the antennas from tabulated system-equivalent flux densities,
and used the average cross-power spectrum in Gate 1 (a smoothed, larger average similar to that shown in Figure\ \ref{proslide}) to estimate the contribution of the source to the spectrum.
Using the normalization condition, we estimated $\tilde \alpha_k$ in regions of the spectrum without pulsar flux.
We used the tabulated average values of $v_0$ and the correlator parameter $n=3$
to determine the constants $A_{X2}$, $B_{X}$, $A_{Y2}$, and $B_{Y}$, for Gates 1 and 6.
Equation \ref{eq:quantizednoise} then provided the noise in the two cases; their ratio is independent of $\Gamma_C$.
We found an expected ratio of 1.23 of the standard deviation of noise in Gate 6 to that in Gate 1.  
This is in approximate agreement with the measured ratio of 1.29, from fits to the histograms shown in Figure\ \ref{empty_dist}.
A more sophisticated calculation might include use of the 
autocorrelation spectra for the two antennas, and quantizer populations measured synchronously with them;
it might also include a spectral model for noise with frequency at each antenna.
% However, such an estimate would still be subject to noise in the autocorrelation and cross-power spectra.

\subsubsection{Noise and Number of Pulses}\label{sec:dutycycle}

The observed distributions of noise are superpositions of underlying distributions.
For example, not all integration times contained the same number of samples.
On the Mopra-Tidbinbilla baseline, the signal was integrated for 2 sec, or 22.4 pulse periods; more precisely,
61\% of the integrations contained 22 pulses, and 39\% contained 23.
Although the data are normalized to the integration time, 
the variance of the noise, parametrized by $b_0$, will be 4\% greater for the integrations with fewer pulses.

\begin{figure}[t]
\epsscale{.80}
% fig 9:
% \plotone{figures/dutycyclecompare.pdf}
\includegraphics*[width=0.98\textwidth]{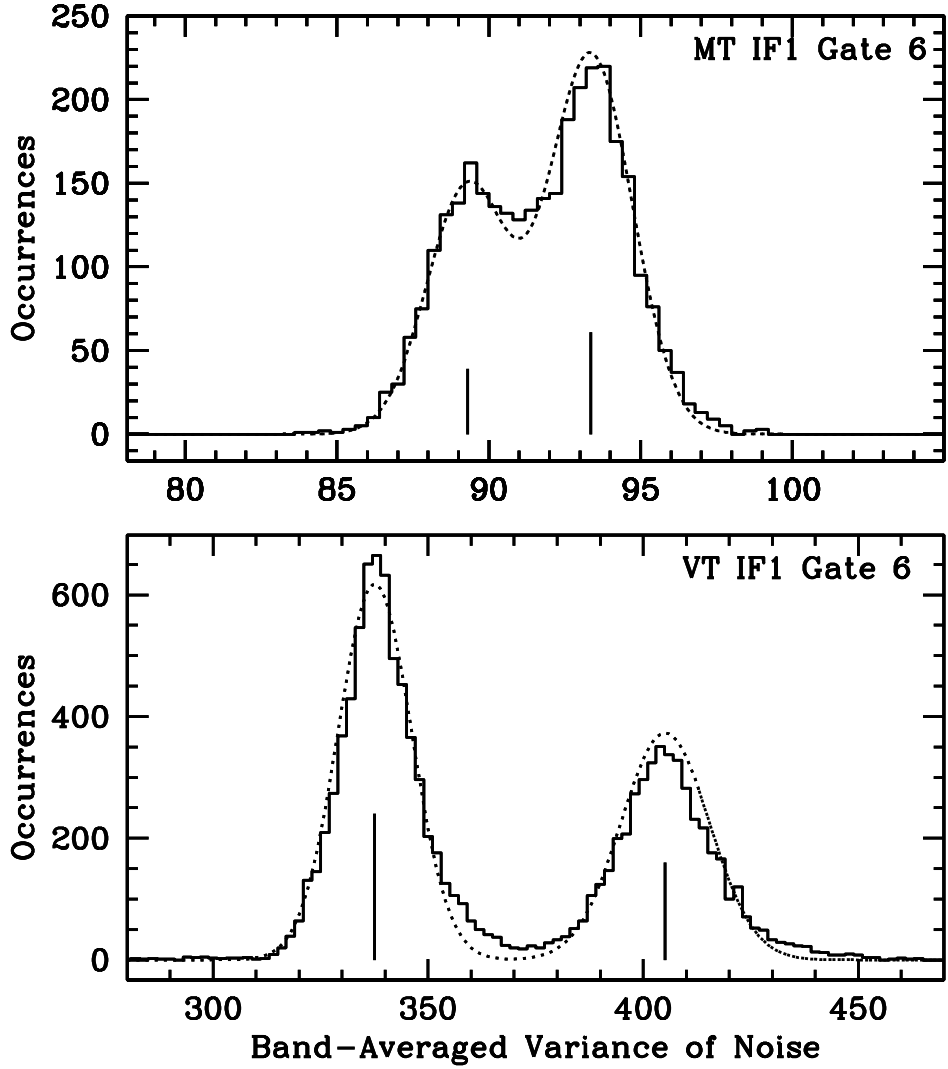}
\figcaption[]{
Distribution of variance of noise in the empty Gate 6,
averaged over the spectrum,
plotted as a histogram.
The vertical lines show the expected horizontal locations of peaks given by the numbers of pulses within an integration time:
22 and 23 for the upper panel (Mopra-Tidbinbilla baseline) and 5 and 6 for the lower panel (VSOP-Tidbinbilla baseline).
Vertical extents of the lines show the expected relative populations of the two peaks.
The dotted lines show a fit for the superposition of 2 Gaussian distributions to the histograms.
\label{fig:duty_cycle_compare}}
\end{figure}

Of course, the number of pulses was the same for all channels within a given sample of the spectrum.
A spectral average provides an estimate of variance.
We demonstrate the effect of integrating over different numbers of pulses in Figure\ \ref{fig:duty_cycle_compare}.
We plot a histogram of the variances over channels 1024 to 2048, for samples from the time span of the data in Section \ref{sec:noise_dist_obs_short}, in IF1.
The histogram shows 2 clear peaks. The centroids of the peaks lie at horizontal positions 
close to the expected ratio of 22:23. 
Their populations are close to the expected ratio of 61:39.
The vertical lines show these positions, constrained to match the overall variance of the data set.
The dotted line shows a simple fit of a model for the sum of two Gaussian distributions to the two peaks,
with the normalizations and locations of the peaks set as for the lines, with net normalization equal to the number of samples,
and the widths of the peaks equal.  
%Thus, the width is the only free parameter in this fit.
% The widths of the Gaussian distributions arise from the finite number of samples, and from variation of the noise across the band, which decreases the effective number of samples.

\subsubsection{Variance: Long Baseline to Spacecraft}\label{sec:empty_gate_dist_variance_long}

% On long baselines noise matches expectation as well.
We analyzed noise for the long baseline from the VSOP spacecraft to
Tidbinbilla, by examining the distribution of noise in individual channels and time samples.
The much smaller number of pulses per 0.5-sec integration on the VSOP-Tidbinbilla baseline 
had a much greater effect on the noise, as Figure\ \ref{fig:duty_cycle_compare} shows.
For this figure, we used IF1, in the empty Gate 6, for the data for Orbit 1.
Comparison of the pulsar period with the integration time indicated that 40\% of samples contain 5 pulses, and 60\%
contain 6 pulses.
Again, the vertical lines under the two peaks show their expected locations, constrained to match the overall variance of the data set.
Locations are in the expected 5:6 ratio, and populations 40:60.
We again show a fit of a model for the sum of two Gaussian distribution to the two peaks.
In this case, the widths of the peaks are proportional to the squares of the variances, as expected for purely statistical contributions to the widths of the peaks.

We found that noise in Gate 6 
from the long baseline was nearly Gaussian,
using an analysis similar to that in Section\ \ref{sec:empty_dist}.
We separated the noise samples into two groups, according to the variance.
The variances of the two groups are 
3.44$\times 10^{-4}$ and 4.12$\times 10^{-4}\ {\rm (correlator\ units)}^2$, respectively.
The variance of the resultant distribution is a weighted sum of the two:
3.71$\times 10^{-4}\ {\rm (correlator\ units)}^2$.
In the empty portion of Gate 1, the statistics are similar, but with smaller resultant variance:
3.10$\times 10^{-4}\ {\rm (correlator\ units)}^2$, for channels 1024 to 2048.
As for the Mopra-Tidbinbilla baseline, the decrease in noise arises from the presence of signal elsewhere in the band.
Using the method discussed in Section\ \ref{sec:empty_gate_dist_variance}, 
we estimated the expected ratio of the standard deviations of noise of in Gate 6 to that in Gate 1 to be 1.19, for the VSOP-Tidbinbilla baseline. 
This expectation is in reasonable agreement with the measured ratio of 1.20.

\subsection{Distribution of Noise with Correlator Lag}\label {sec:not_wrapping}

Variances and covariances completely characterize Gaussian noise.
Covariance of noise in different spectral channels can arise from quantization \citep{gwi06}, 
from source variability on short timescales
\citep{gwi11a,gwi11b}, and from pulsar gating, as described in the following section.
In this section, we calculate the covariance for noise in an empty gate and compare with observations.
We then discuss how, for gates containing signal, the covarance introduced by source variability masks that from pulsar gating. 

\subsubsection{Pulsar Gate, Without Wrap}\label{sec:empty_corr_theory}

Pulsar gating can introduce spectral correlations.
Indeed, any temporal modulation is predicted to introduce spectral correlation \citep{gwi11a}.
%We investigate the presence of these correlations in our observations.
%
For our observations, the maximum lag correlated approaches the width of a pulse gate.
Therefore, at large lag, we obtain fewer measurements of the correlation function than at small lag. 
%Normalization by the number of samples
%ensures that the average correlation function equals that which would be obtained
%with uniform sampling; however, the noise is greater at larger lag 
%and thereby introduces correlations
%of the noise between channels in the Fourier-conjugate spectral domain.
%Alternatively, if the correlation function ``wraps'' then the noise is uniform;
%this wrapping amounts to the assumption that for a sample of length $N_{\rm obs}$,
%the correlation function at lag $\tau$ is equivalent to that at lag $N_{\rm obs}-\tau$.
%\citet{gwi06} and \citet{gwi11a} made this assumption.
%Most FX correlators satisfy this assumption,
%but most XF correlators do not.
%
The covariance of noise between two spectral channels
is given by
Fourier transform of the product of elements of the
correlation function.
% as argued in Section\ 2.3 of \citet{gwi06}.
The Fourier transform of two copies of Equation\ \ref{eq:rtau_def} leads to an expression similar to Equation\ 22 of \citet{gwi06}:
\begin{equation}
\langle \tilde r_k \tilde r_{k+\ell}^*\rangle - \langle \tilde r_k\rangle\langle \tilde r_{k+\ell}^*\rangle= 
\sum_{\upsilon,\mu=-N}^{N-1}
{\frac{1}{(N_{\rm obs}-|\upsilon+\mu|)(N_{\rm obs}-|\upsilon|)}}
\sum_{n,m}
e^{\left[ i {\frac{2\pi}{2 N}} \left( k \mu + (k-\ell) \upsilon\right)\right] }
\alpha_{n-m} \alpha_{-(n-m)+\mu}.
\label{eq:spectral_correlation}
\end{equation}
Note that, in our case, the limits of the sums over $n$ and $m$ 
depend on both $\upsilon$ and $\mu$.
This is because the
correlation functions $r_{\tau}, r_{\nu}^*$ are not averaged over all of the samples
at both stations.
%Therefore, the argument of the sum in Equation\ \ref{eq:spectral_correlation} depends on $\upsilon$ and $\mu$, 
%and the sum over $\upsilon$ does not yield zero for all $k\neq \ell$, as it did in \citet{gwi06}.
%
The correlation of noise between spectral channels given by Equation\ \ref{eq:spectral_correlation} depends on the 
specific form of the 
autocorrelation function, $\alpha_\tau$.
However, the autocorrelation function is always maximum at $\alpha_0 = 1$ for lag $\tau=0$, 
and usually falls rapidly to zero for larger lags $\tau$.
%For spectrally-flat signal and noise,
%the autocorrelation function is nonzero only for $\tau=0$.
%We explore this case in the following section.

\subsubsection{Correlation of Noise: White Signal and White Noise}\label{sec:empty_corr_theory_nosignal}

We suppose in this section that the original time series is spectrally-flat or ``white'' noise \citep{Pap91}.
Consequently, the correlation function is zero except at the central lag:
$\alpha_0 =1$, and $\alpha_\upsilon=0$ for $\upsilon\neq 1$.
We then find:
\begin{equation}
\langle r_{\upsilon} r_{\upsilon}^* \rangle - \langle r_{\upsilon}\rangle \langle r_{\upsilon}^* \rangle
={\frac{1}{N_{\rm obs}-|\upsilon|}}.
\label{statistical_lag_noise_white}
\end{equation}
%As expected, the noise in the correlation function depends on the lag.
% This expression simply reflects the fact that we have only $N_{\rm obs}-|\upsilon|$ samples of the correlation function at lag $\upsilon$.
Because we have normalized the correlation function and the signal is white,
the mean square noise is just the reciprocal of the number of samples.

In the spectral domain, this variation of the number of samples leads to correlation.
In Equation\ \ref{eq:spectral_correlation},
the product of $\alpha$'s is zero unless $m=n$ and $\mu=0$.
Performing the remaining sums, we find for the correlation between channels:
\begin{equation}
\langle \tilde r_k \tilde r_{k+\ell}^*\rangle - \langle \tilde r_k\rangle\langle \tilde r_{k+\ell}^*\rangle= 
\sum_{\upsilon=-N}^{N-1}e^{\left[ i {\frac{2\pi}{2 N}} (k-\ell) \upsilon \right] }
{\frac{1}{N_{\rm obs}-|\upsilon|}}.
\end{equation}
% Thus, an uneven noise distribution in the correlation domain introduces correlations of noise in the spectral domain.
This correlation is largest for close spectral channels ($\ell\ll 2N$)
because the noise in the correlation function is largest for large lag ($\upsilon\rightarrow N$).

\subsubsection{Observed Correlation of Noise in an Empty Gate}\label{sec:empty_gate_corr}

\begin{figure}[t]
\epsscale{.80}
% fig 10:
% \plotone{figures/noise6_spect_dingledouble.pdf}
\includegraphics*[width=0.98\textwidth]{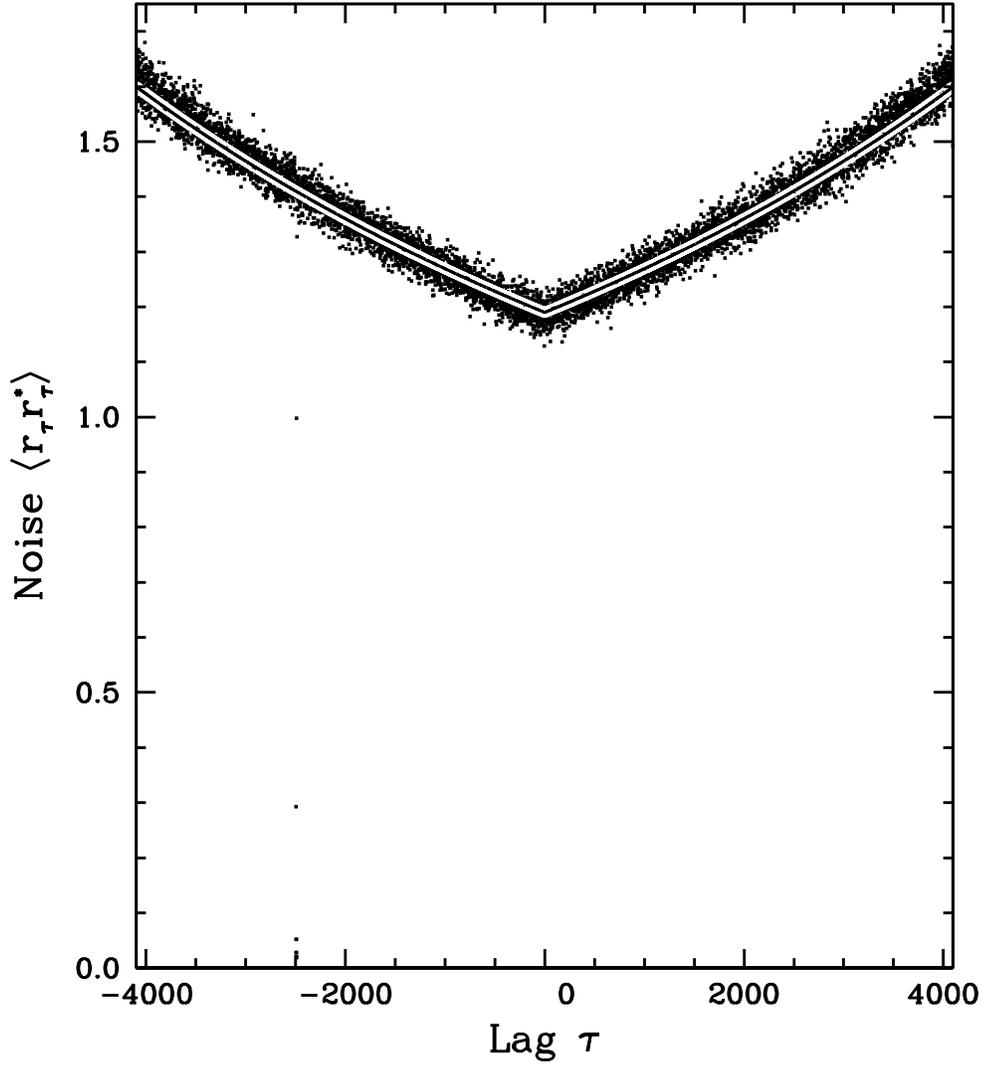}
\figcaption[]{
Correlation function of noise in empty Gate 6 on the MT baseline, plotted with lag $\tau$.  
Points show measured mean-square correlation function.  Solid line shows theoretical curve for effects of pulse gating.
A single parameter, the overall scale, has been fit to match theory to observation.
\label{fig:corr_empty}}
\end{figure}

We compared observations on the Mopra--Tidbinbilla baseline with the theoretical prediction
of Equation\ \ref{statistical_lag_noise_white},
and find quantitative agreement.
Figure\ \ref{fig:corr_empty} shows the results.
To make the figure,
we re-transformed the spectrum for each time interval, and
found the mean square correlation function
$\langle r_{\tau} r_{\tau}^* \rangle$.
The data used to make Figure \ref{fig:corr_empty} are from Gate 6, IF2, of the Tidbinbilla--Mopra baseline,
from 14:15 to 15:41 UT.  We used the full recorded bandwidth.
The downward-pointing spike at $\tau=5000$
apparently results from interference.
The solid curve shows the form predicted by Equation\ \ref{statistical_lag_noise_white},
as expected for
$N_{\rm obs}=16{,}000$\ complex samples, for the 1-msec length of the pulsar gate and our 16\ MHz bandwidth.

Figure\ \ref{fig:corr_empty} shows less noise at small lag,
as Equation\ \ref{statistical_lag_noise_white} suggests.
% Note that, in this case, the length of the gate sets $N_{\rm obs}$, rather than the total integration time.
% The model would fit slightly better if the number of samples $N_{\rm obs}$ were smaller;  truncation effects in the correlator may be responsible for the discrepancy.
%
%Figure\ \ref{fig:corr_empty} shows that the noise of the correlation function is greater for larger lags,
%as expected.
%Those large lags
%correspond in the Fourier-conjugate spectral
%domain to many oscillations across the spectral range.
%Thus, noise with rapid spectral oscillations is over-represented,
%relative to uncorrelated white noise,
%and adjacent spectral channels will be anti-correlated.
%% One can determine the degree of correlation by 
From Fourier transform of
the theoretical curve shown in the figure,
we find that noise in the spectrum $\tilde r_k$ is anticorrelated in adjacent channels
by approximately $-6$\%, expressed as normalized correlation.
This correlation falls off rapidly, however: it is only $+$0.4\% in the second channel,
and $-0.7$\% in the third, and so on.

% XXXX this is the first time that this correlation has been calculated: 9a

\subsubsection{Correlation of Noise: Signal Present}\label{sec:empty_corr_theory_withsignal}

If the spectrum is not ``white'', then Equation\ \ref{eq:spectral_correlation} still holds, and $\alpha_0 = 1$.
%Although the noise and correlations in the conjugate spectral domain 
%depend on the exact form of $\alpha_{\tau}$, we can anticipate some of their features.
For typical spectra, we expect that correlations away from the zero-lag will be small: $|\alpha_{\tau}|\ll 1$, for $\tau\neq 0$.

If all of the features of the spectrum are fully spectrally resolved, 
then any nonzero values of $\alpha_\tau$ will be concentrated in a range close to $\tau=0$,
and the effect of the reduced denominator on the right-hand side of Equation\ \ref{eq:spectral_correlation}
will tend to increase the noise at large lag $\tau$.
On the other hand, if the spectrum is not fully resolved, 
some nonzero values of $\alpha_{\tau}$ will be missing from the sum,
which will tend to decrease the noise at large lags.
Both effects are larger for higher lags $\tau$ and, thus, will
tend to introduce small-lag correlations in the conjugate spectral domain.
The first will introduce negative correlations, the second positive correlations.
For well-resolved spectra,
the first effect will predominate,
and will likely produce anticorrelations comparable to those estimated for white spectra and noise.

\section{CONCLUSIONS}\label{sec:conclusions}

\subsection{Summary of Results}\label{sec:summary}

We compare theoretical predictions for the distribution of noise 
for cross-power spectra with observations of
a scintillating pulsar, the Vela pulsar.
We describe observations made with Earth-based VLBI baselines, and with baselines from an orbiting spacecraft to Earth.  
These observations extend previous studies \citep{gwi11a} to the regime of high signal-to-noise ratio and large variations in interferometric phase.

In Section\ \ref{sec:theory_bckgnd}, we argue that, in the presence of signal, 
noise on a short baseline should be drawn from an elliptical Gaussian distribution 
in the complex plane.
The theory was previously presented in \citet{gwi06} and \citet{gwi11a}.
The major axis of the distribution is aligned with the direction of the signal.
The variance along the minor axis depends linearly on signal strength;
the variance along the major axis has the same linear dependence, plus a quadratic term.
At zero signal, the major and minor axes are equal and the distribution of noise is a circular Gaussian,
as for a gate or spectral region empty of signal.

We test this theory with observations on the baselines from Mopra, Hartebeesthoek, and
the VSOP spacecraft to Tidbinbilla in Section\ \ref{sec:noise_dist_obs}.
We estimate noise by comparing samples within the characteristic scales of the scintillation,
and binning their differences by average interferometric visibility. 
We find that the distribution of noise closely follows the expected elliptical Gaussian form for each visibility, and the scaling with visibility of the major and minor axes corresponds to the quadratic and linear noise polynomials, respectively. The quadratic coefficient accurately reflects the number of samples and the contribution of intrinsic amplitude variations; the constant coefficient agrees with that estimated from empty portions of the spectrum
for the Mopra-Tidbinbilla and Hartebeesthoek-Tidbinbilla baselines.

In Section\ \ref{sec::Correlator_Effects}, we demonstrate that quantization, gating, and integration each affect the properties of the noise. One interesting consequence is that the noise in the presence of signal is less than that in a completely empty spectrum -- a result of the combination of quantization and spectral variations. Also, pulsar gating leaves fewer samples, thus larger noise, at larger lags; this effect incurs correlations in the spectral noise. In principle, complete knowledge of the quantizer levels for each integration period, and of the autocorrelation functions at the two antennas, allows calibration of these effects.% \citep[see][]{jen98}. 
Alternatively, recording the signal with many quantizer levels increases the dynamic range, and reduces effects of variation in quantizer levels. Flexible software correlators, such as the DiFX correlator \citep{del07}, can control these artifacts, while Nyquist-sampled spectra of individual pulses obviate the difficulties in characterizing inhomogeneities within the integration \citep{Joh12a}.

\subsection{Self-Noise for an Interferometer}

We present observations for a scintillating pulsar, but the effects of self-noise hold for any 
interferometric observation. A careful evaluation of these effects is essential for a priori estimates of the accuracy of pulse-timing and spectroscopy using single-dish observations, and for scintillation studies and astrometry using interferometry. 

Many telescopes now under construction or being planned, such as LOFAR, ASKAP, SKA, and so on will operate as interferometers,
with many baselines among many antennas of a particular design. Each baseline will have the distribution of noise 
we describe above in Section\ \ref{sec:theory_bckgnd},
and as we observe for the Vela pulsar. Because each telescope in a large array receives precisely the same noiselike signal from the source, increases in the number of antennas and averaging of many baselines
do not change self-noise, when expressed in terms of flux density. 
However, averages over many baselines do decrease the background noise.
Stated in terms of the notation introduced in Section\ \ref{sec:theory_bckgnd}, increasing the number
of identical antennas $N_A$ reduces $b_0$  as $N_A^{-2}$ and $b_1$  as $N_A^{-1}$
but does not change $b_2$.
For true ``tied-array'' operation,
where electric fields from all antennas are phased and summed before correlation,
statistics are those of a single dish, \citep[][Equation\ 11]{gwi11a}.
As the number of antennas become larger, self-noise becomes more important.
When the source dominates the system temperature, further improvements
demand more samples $N_{\rm obs}$, as produced by wider bandwidth or longer integration time: a greater aperture does not provide more accuracy.
Our expression Equation\ \ref{eq:noise_perp_para} generalizes this result to interferometry.

Astrometry depends on measurements of interferometric phase. 
Equations\ \ref{eq:noise_perp_para} and \ref{eq:dicke_complex}
show that the maximum attainable phase accuracy is $\delta\phi \approx \sigma_{\perp}/s \approx \sqrt{n/(N_{\rm obs} s)}$,
or the inverse square root of the signal-to-noise ratio $s/n$, divided by $\sqrt{N_{\rm obs}}$.
Similarly, the maximum attainable accuracy in measurement of flux density 
by an interferometer, or a large single dish, is approximately the flux density of the source divided by $\sqrt{N_{\rm obs}}$.
Likewise, the maximum attainable accuracy in pulsar timing is approximately the width of the narrowest feature in the profile,
divided by the signal-to-noise ratio and by $\sqrt{N_{\rm obs}}$.  However, when self-noise is the limiting factor,
the maximum attainable accuracy is simply the width of that feature, divided by $\sqrt{N_{\rm obs}}$.

\acknowledgments

I gratefully acknowledge the VSOP Project, which is led by the Japanese 
Institute of Space and Astronautical Science in cooperation with many 
organizations and radio telescopes around the world.
I am grateful 
to the DRAO for supporting this work with extensive
correlator time.
We thank the U.S. National Science Foundation for financial support for this work
(AST 97-31584 and AST-1008865).

%\appendix
%\section{Useful Facts for Spectra}\label{useful_spectral_facts}

\newpage

\newpage

\newpage

\newpage

\newpage

\newpage
\newpage

\newpage

\newpage

\newpage

\newpage

\clearpage

\begin{deluxetable}{lcccc}
\tablenum{1}
\tablecolumns{5}
\tablewidth{0pc}
\tablecaption{Quantizer Levels: $\pm v_0$}
% 
% /Users/carlgwinn/Desktop/VSOP_True/noise_adiff/levels/correlator_stats
% levels.mac
%  XXX These are indeed std dev, not variances CRG 25 May 2012
\tablehead{
& 
\multicolumn{2}{c}{On Pulse:}&
\multicolumn{2}{c}{Off Pulse:}\\
&
\multicolumn{2}{c}{Gate 1, IF1}&
\multicolumn{2}{c}{Gate 6, IF1}\\
\colhead{Station}& 
\colhead{Average}&
\colhead{Std. Dev.}&
\colhead{Average}&
\colhead{Std. Dev.}
}
\startdata
% Hartebeesthoek    &    0.928259  & 0.006961&  0.944029  & 0.006926 \\
% Mopra                   &    0.933236  & 0.004773& 0.943795   & 0.004841 \\
% Tidbinbilla             &    0.843062  & 0.017374& 0.947405   & 0.004375 \\
% VSOP Spacecraft  &   0.922964  & 0.014065 &  0.923391  & 0.014247 \\
Hartebeesthoek    &    0.928  & 0.007\phantom{$^{b}$}&  0.944  & 0.007\phantom{$^{b}$} \\
Mopra                   &    0.933  & 0.005\phantom{$^{b}$}& 0.944   & 0.005\phantom{$^{b}$} \\
Tidbinbilla             &    0.843  & 0.017$^{a}$& 0.947   & 0.004\phantom{$^{b}$} \\
VSOP Spacecraft  &   0.923  & 0.014$^{b}$ &  0.923  & 0.014$^{b}$ \\
\enddata
\tablenotetext{a} {Standard deviation reflects pulse-to-pulse variations
in intensity.}
\tablenotetext{b} {Standard deviation reflects primarily a slow drift over
the time span of observations.}
\label{table_quantizer_v0}
\end{deluxetable}

\clearpage
\end{document}